\documentclass[10pt,twocolumn,journal]{IEEEtran}
\usepackage{latexsym}
\usepackage{amsfonts}
\usepackage{amsbsy}
\usepackage{amsmath,amssymb,amsfonts,bm,amsthm}
\usepackage{amsthm}
\usepackage{times}
\usepackage{graphicx}
\usepackage{enumerate}
\usepackage[usenames]{color}
\usepackage[dvips]{pstcol}
\usepackage{epstopdf}
\usepackage{cite}
\usepackage{amsmath}
\usepackage{cases}
\usepackage{amssymb}
\usepackage{amsfonts}
\usepackage{graphicx}
\usepackage{epsfig}
\usepackage{psfrag}
\usepackage{xcolor}
\usepackage{amsfonts, bm}
\usepackage{epstopdf}
\usepackage{cite}
\usepackage{color}
\usepackage{xcolor}
\usepackage{verbatim}
\usepackage{subfig}
\usepackage{algorithm}
\usepackage{algorithmic}
\usepackage[T1]{fontenc}
\usepackage{colortbl}
\newtheorem{remark}{Remark}

\input epsf
\begin{document}
\title{\huge Latency Minimization for mmWave D2D Mobile Edge Computing Systems: Joint Task Allocation and Hybrid Beamforming Design}

%%%%%%%%%%%%%%%%%%%%%%%%%%%%%%%%%%%%%%%%

\author{Yanzhen Liu, Yunlong Cai,  An Liu, Minjian Zhao, and Lajos Hanzo
\thanks{
Copyright (c) 2015 IEEE. Personal use of this material is permitted. However, permission to use this material for any other purposes must be obtained from the IEEE by sending a request to pubs-permissions@ieee.org.

The work of Y. Cai was supported in part by  the National Natural Science
Foundation of China under Grants 61971376 and 61831004, and the Zhejiang
Provincial Natural Science Foundation for Distinguished Young Scholars
under Grant LR19F010002.
L. Hanzo would like to acknowledge the financial support of the
Engineering and Physical Sciences Research Council projects EP/P034284/1
and EP/P003990/1 (COALESCE) as well as of the European Research
Council's Advanced Fellow Grant QuantCom (Grant No. 789028). (Corresponding author: Yunlong Cai.)

Y. Liu, Y. Cai, A. Liu and M. Zhao are with the College of Information Science and Electronic Engineering, Zhejiang University, Hangzhou 310027, China (e-mail: yanzliu@zju.edu.cn; ylcai@zju.edu.cn; anliu@zju.edu.cn; mjzhao@zju.edu.cn).

L. Hanzo is with the Department of ECS, University of Southampton, UK (e-mail: lh@ecs.soton.ac.uk).
}
}

\maketitle
\vspace{-3.3em}
\begin{abstract}
Mobile edge computing (MEC) and millimeter wave (mmWave)
communications are capable of significantly reducing the network's
delay and enhancing its capacity. In this paper we investigate a mmWave and
device-to-device (D2D) assisted MEC system, in which user A carries out some computational tasks and shares the results with user B with the aid of
a base station (BS). We assume partial offloading model and the task
can be partitioned into two portions: the first part is computed
locally at user A, while the second part is transmitted to the BS and
computed by the MEC server. The computational results are then sent to
user B through a D2D link and via the link from the BS to user B,
respectively. To support computation offloading, both
the users and the BS are equipped with multiple antennas and employ
analog and digital (A/D) hybrid beamforming. {\color{black}Moreover, we propose a novel two-timescale joint hybrid beamforming and task allocation algorithm to reduce the system latency whilst cut down the required signaling overhead. Specifically, the high-dimensional analog beamforming matrices are updated in a frame-based manner based on the channel state information (CSI) samples,  where each frame consists of a number of time slots, while the low-dimensional digital beamforming matrices and the offloading ratio are optimized more frequently relied on the low-dimensional effective channel matrices in each time slot. A stochastic successive convex approximation (SSCA) based algorithm is developed to design the long-term analog beamforming matrices. As for the short-term variables, the digital beamforming matrices are optimized relying on the innovative penalty-concave convex procedure (penalty-CCCP) for handling the mmWave non-linear transmit power constraint, and the offloading ratio can be obtained via the derived closed-form solution. Simulation results verify the effectiveness of the proposed algorithm by comparing the benchmarks.}
\end{abstract}
%\vspace{-1.3em}

\IEEEpeerreviewmaketitle

\section{Introduction}
\label{sec:intro}
Given the rapid growth of computational-intensive mobile applications such as virtual
reality (VR)~\cite{VR}, augmented reality (AR)~\cite{AR}, automatic driving~\cite{Autodriving}, and face recognition~\cite{facerecognition}, conventional remote cloud computing centers tend to struggle in
meeting the stringent latency requirements of next-generation wireless
systems~\cite{MEC_background1}.  Mobile edge computing (MEC) - which supports servers at the base station (BS) of
cellular networks - has emerged as a promising solution~\cite{MEC_background2,MEC_intro}. Thanks
to the proximity of the mobile devices to the server,  users can directly offload the computational-intensive tasks to the edge server without passing the back-haul networks, which significantly reduces the end-to-end delay and the network burden\cite{MEC_bi1,MEC_bi2,MEC_bi3,MEC_bi4,MEC_bi5,MEC_pa1,MEC_pa2,MEC_pa3,MEC_pa4,MEC_pa5}. Specifically, the works in~\cite{MEC_bi1,MEC_bi2,MEC_bi3,MEC_bi4,MEC_bi5} considered the binary offloading in MEC systems. The authors of~\cite{MEC_bi1} studied an energy efficient binary offloading problem and designed optimal scheduling policies for both the mobile execution and cloud execution. An MEC system combined with energy harvesting techniques has been investigated in~\cite{MEC_bi2,MEC_bi3}. In~\cite{MEC_bi4}, the authors proposed a general framework for offloading tasks from a single user to multiple access points. Moreover, the authors of~\cite{MEC_bi5}  investigated a joint design problem of the computation offloading decision, the resource allocation, and the content caching strategy. The partial offloading schemes have been proposed to further improve the performance of MEC systems ~\cite{MEC_pa1,MEC_pa2,MEC_pa3,MEC_pa4,MEC_pa5}. In~\cite{MEC_pa1}, an optimal resource allocation
scheme has been proposed for a multi-user MEC system based on time-division multiple access (TDMA) and orthogonal frequency-division multiple access (OFDMA), respectively. The authors
of~\cite{MEC_pa2} studied
a multi-user TDMA partial offloading MEC system, and derived the optimal solution to the delay
minimization problem. By taking user cooperation into consideration, the authors of~\cite{MEC_pa3} investigated an energy-efficient problem for both binary offloading and partial offloading. To improve edge cloud efficiency with limited communication and computation capacities, the collaboration between cloud computing and edge computing was studied in \cite{MEC_pa4,MEC_pa5}. Furthermore, the authors of \cite{MEC_IRS1,MEC_IRS2,MEC_IRS3} investigated the intelligent reflecting surface (IRS) assisted MEC systems to improve the network efficiency.

However, MEC needs frequent data exchange between the mobile devices and the edge server, which requires a large communication capacity of the radio access network. Taking the $360$-degree immersive VR as an example, even under the $265$ HEVC $1:600$ video compression rate, a bit rate of up to 1 Gbps~\cite{VR_rate} is needed to match the $2\times 64$ million pixel human-eye accuracy, which is challenging for the current 5th generation (5G) mobile communication technology~\cite{VR_rate2}. Therefore, it is necessary to further enhance the system capacity for beyond 5G MEC systems. The millimeter wave (mmWave) and device-to-device (D2D) communications are exactly two promising techniques. MmWave has tremendous spectral resources and can achieve multi-gigabit transmission capacity. Moreover, at this short
wavelength it is possible to integrate a large number of antenna
elements in a compact space \cite{mmWave_survey}, thus achieving significant beamforming gain. D2D communication supports multiplex of the cellular spectrum, which allows mobile devices in proximity to communicate directly. It features high data rate, low latency, and high throughput~\cite{D2D_feature}, which fits the communication needs of MEC systems well.  As a result, a number of solutions applying D2D techniques to MEC systems have been proposed  to enable direct data transmission and computational resource sharing~\cite{MECD2D_1,MECD2D_2,MECD2D_3,MECD2D_stochastic}.

To the best of our knowledge,  the mmWave and D2D assisted MEC has not been well investigated in the literature. Although the D2D-aided MEC systems have been studied in the aforementioned works~\cite{MECD2D_1,MECD2D_2,MECD2D_3,MECD2D_stochastic}, they assumed sub-6GHZ band and the tremendous mmWave spectral resource has not been considered to further enhance the capacity. Moreover, despite the tremendous benefits brought by the integration of mmWave, D2D and MEC, there are  more challenges compared with existing works. To elaborate,  1) The challenges of the physical layer
signal processing incurred in the mmWave frequency band, such as hybrid analog and digital (A/D) beamforming~\cite{mmWave_AO,mmWave_OMP,mmWave_PDD1,mmWave_PDD2,mmWave_heuristic,channelmatching}, associated non-linear power consumption model, CSI acquisition etc.
2) The design of practical protocols to avoid the occurrence of transmission collisions that may happen in
simultaneous uplink/downlink and D2D transmissions. 3) The design of efficient algorithms to solve the challenging non-convex optimization problems.

Hence, to fill this research blank and tackle the above challenges, we investigate a mmWave and D2D assisted MEC system, where
user A processes the computational tasks to be solved and then shares
the results with user B with the aid of a BS. The investigated model is general and its typical application scenarios include vehicle to vehicle communication~\cite{V2V}, VR/AR gaming~\cite{D2D_VR}, and ultra high definition video transmission~\cite{D2D_HD}. We assume partial offloading model as in~\cite{MEC_pa1,MEC_pa2,MEC_pa3,MEC_pa4,MEC_pa5}, i.e., the task of user A can
be partitioned into two portions: the first part is computed locally
at user A, while the second part is transmitted to the BS and computed
by the MEC server. The computation results are then sent to user B
through a D2D link and the link from the BS to user B, respectively. In order to support computation offloading,
both the users and the BS are equipped with multiple antennas and
employ A/D hybrid beamforming. \textcolor{black}{However, directly solving it by using the single-timescale algorithm requires very high complexity and a large amount of CSI feedback. Thus, we propose a novel two-timescale joint hybrid beamforming and task allocation algorithm to reduce the system latency whilst cut down the required signaling overhead. Specifically, the high-dimensional analog beamforming matrices are updated in a frame-based manner based on the channel state information (CSI) samples, where each frame consists of multiple time slots, while the low-dimensional digital beamforming matrices and the offloading ratio are optimized more frequently relied on the low-dimensional effective channel matrices in each time slot.  We respectively formulate a long-term weighted ergodic channel capacity maximization problem and a short-term latency minimization problem for practical design.} Our main contributions are summarized as follows:
\begin{itemize}
	{\black
	\item We study a novel scenario that combines MEC with mmWave and D2D to significantly reduce the delay. We consider the raw data and result transmission in the uplink, the downlink, and the D2D link in details and make practical protocols to avoid collisions.
	\item For the long-term weighted ergodic channel capacity maximization problem, a stochastic successive convex approximation (SSCA) based algorithm is developed for designing the analog beamforming matrices, which employs surrogate functions to approximate the original problem and  converges to a stationary feasible solution.
	\item Regarding the design of the digital beamforming matrices and offloading ratio, we equivalently decompose the short-term latency minimization problem into several decoupled subproblems. For the subproblems w.r.t. the digital beamforming matrices, an efficient penalty-CCCP based algorithm is proposed to tackle the nonlinear mmWave transmit power constraints. We also develop a low-complexity heuristic algorithm to design the digital beamforming matrices for performance-complexity trade-off.}
	%\item \textcolor{blue}{For the digital beamforming matrices design, we consider the practical mmWave power consumption model.} To tackle the resultant subproblems of digital beamforming matrices, .
	%This algorithm is a double-loop one where in the inner loop we approximate the original problem using the convex-concave procedure (CCCP)~\cite{ccp} and update the variables in a block coordinate descent (BCD) fashion, while in the outer loop we adjust the penalty parameter.
	\item  The closed-form expressions of offloading ratio are derived based on classified discussion.  We compare our proposed joint design algorithm of the task allocation and hybrid beamforming with the conventional algorithms in the simulation. The results verify the effectiveness of our proposed joint design algorithm.
\end{itemize}

The paper is structured as follows. Section \ref{Section2:system} describes the system model. \textcolor{black}{Section \ref{Section3:problem} formulates the two-timescale problem under investigation. The long-term analog beamforming design problem is solved
in Section \ref{Section4:longterm} while the solutions to the design of the short-term digital beamforming matrices and the optimal offloading ratio are given in Section \ref{Section5:shortterm}}. Section \ref{Section6:simulation} presents the simulation results and Section \ref{Section7:conclusion} concludes this paper.

\emph{Notations:}
%\textcolor{black}{The notations used in this paper are summarized in Table~\ref{table1}.}
Scalars, vectors and matrices are denoted by lower case, boldface lower case and boldface upper case letters, respectively. $\mathbf{I}$ represents an identity matrix and $\mathbf{0}$ denotes an all-zero  matrix.
For a  matrix $\mathbf{A}$, ${{\mathbf{A}}^T}$, $\textrm{conj}(\mathbf{A})$, ${{\mathbf{A}}^H}$, $\mathbf{A}^{\dagger}$ and $\|\mathbf{A}\|$ denote its transpose, conjugate, conjugate transpose, Moore-Penrose inverse and Frobenius norm, respectively. For a square matrix $\mathbf{A}$, $\textrm{Tr} \{\mathbf{A}\}$ and $\mathbf{A}^{-1}$ denotes its trace and inverse, respectively, while ${\mathbf{A}} \succeq {\mathbf{0}}~({\mathbf{A}} \preceq {\mathbf{0}})$ means that $\mathbf{A}$ is  positive (negative) semi-definite.
For a vector $\mathbf{a}$, $\|\mathbf{a}\|$ represents its Euclidean norm.
$\Re\{\cdot\}$ ($\Im\{\cdot\}$) denotes the real  (imaginary) part of a variable.
$|  \cdot  |$ denotes  the absolute value of a complex scalar.
${\mathbb{C}^{m \times n}}\;({\mathbb{R}^{m \times n}})$ denotes the space of ${m \times n}$ complex (real) matrices. $\angle$ denotes the angle operator.
%\begin{table}[htbp]
%	\centering
%	\caption{List of Notations}
%	\label{table1}
%	\begin{tabular}{|c|c|c|c|}
%		\hline
%		%& & & \\[-6pt]
%		Symbol&Representation&Symbol&Representation \\
%		\hline
%		%& & & \\[-6pt]
%		$a$&(complex)scalar&$\mathbf{I}$ ($\mathbf{0}$)&identity (zero) matrix \\
%		\hline
%		%& & & \\[-6pt]
%		$\mathbf{a}$&vector&$\textrm{Tr} \{\mathbf{A}\}$ &trace \\
%		\hline
%        %& & & \\[-6pt]
%        $\mathbf{A}$&matrix&$\mathbf{A}^{-1}$ &inverse \\		
%		\hline
%		%& & & \\[-6pt]
%		$\mathbf{A}^T$&transpose&${\mathbf{A}} \succeq {\mathbf{0}}~({\mathbf{A}} \preceq {\mathbf{0}})$ & positive (negative) semi-definite \\
%		\hline
%		$\text{conj}(\mathbf{A})$&conjugate&$\Re\{\cdot\}$ ($\Im\{\cdot\}$) &real (imaginary) part of a variable \\
%		\hline
%		$\mathbf{A}^H$&conjugate transpose&$|  \cdot  |$ &absolute value of a complex scalar \\
%		\hline
%		$\mathbf{A}^{\dagger}$&Moore-Penrose inverse&${\mathbb{C}^{m \times n}}\;({\mathbb{R}^{m \times n}})$ &space of ${m \times n}$ complex (real) matrices \\
%		\hline
%		$\|\mathbf{A}\|$&Frobenius norm&$\angle$ &angle operator \\
%		\hline
%		$\otimes$&Kronecker product&$\circ$ &Hadamard product \\
%		\hline
%	\end{tabular}
%\end{table}

\section{System Model}
\label{Section2:system}
\begin{figure}[t]
	\centerline{\includegraphics[width=3.3in]{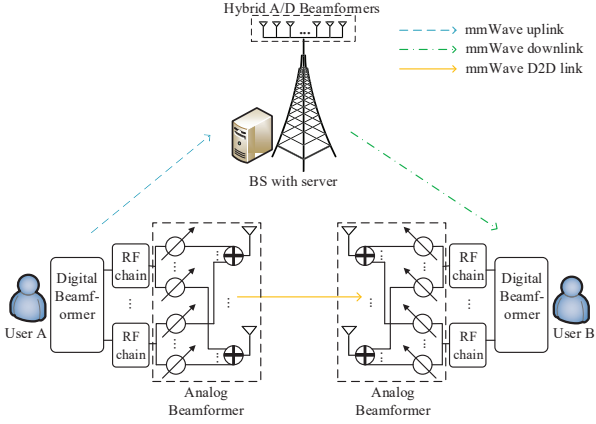}}
	\caption{mmWave D2D MEC system.}
	\label{figure1} \vspace{-1em}
\end{figure}
In this section, we introduce the investigated system model. As shown in Fig.~\ref{figure1}, we consider a system consisting of user A, user B and a BS with a MEC server. User A aims to process computation tasks and share the results with user B with the aid of the BS. We assume partial offloading model as~\textcolor{black}{as \cite{MEC_pa1,MEC_pa2,MEC_pa3,MEC_pa4,MEC_pa5}}, i.e., user A has a total of $L$ bits task to be processed, and this task can be divided into two parts: $\rho L$ bits and $(1-\rho)L$ bits, where $\rho$ denotes the offloading ratio. The first part is transmitted to the BS and computed by the MEC server, and the second part is computed at the local CPU of user A. The computational results are transmitted to user B through the D2D link and the downlink (between the BS and user B), respectively.\footnote{It is worth mentioning that unlike the works in \cite{MECD2D_3} where the authors utilize the computation resource of both the MEC server and the D2D users, we do not use the computation resource of user B because transmitting the raw data is time-consuming while the computing capacity at user B has no advantages over that of the BS.} Both the users and the BS are equipped with A/D hybrid beamformers and work in mmWave band (The hybrid beamforming architecture of the BS is not plotted here since it is similar with that of the users).

\subsection{Computation model}
\textcolor{black}{In this paper, we adopt a general compression model and denote the computational results as $\alpha L$ bits, where $0\leq \alpha \leq 1$ denotes the compression ratio for the computation task and can be chosen as different values based on the category of the task and the adopted algorithm \cite{data_compression_model}.} Defining
$K_L\triangleq\frac{L}{F_L}$, $K_E\triangleq\frac{L}{F_E}$,
$K_{1}\triangleq\frac{L}{R_{1}}$, $K_{2}\triangleq\frac{\alpha
	L}{R_{2}}$ and $K_{3}\triangleq\frac{\alpha L}{R_{3}}$ for
convenience of notation, where $F_L$ and $F_E$ stand for the local computing capacity and the edge computing capacity (computing capacity of the MEC server), respectively, and $R_{1}$, $R_{2}$ and $R_{3}$ represent the transmission rates of the uplink (from user A to the BS), downlink and D2D link, respectively. We express different delays as follows,

\begin{itemize}
	\item The local computing time:  $T_L^c=(1-\rho)K_L$.
	\item The computing time at the MEC server:  $T_E^c=\rho K_E$.
	\item The offloading time from user A to the BS: $T_{up}^t=\rho K_1$.
	\item The delay for transmitting the computational result from the BS to user B: $T_{down}^t=\rho K_2$.
	\item The delay for transmitting the computational result from user A to user B: $T_{D2D}^t=(1-\rho)K_3$.
\end{itemize}

\begin{figure}[t]
	\centerline{\includegraphics[width=3.8in]{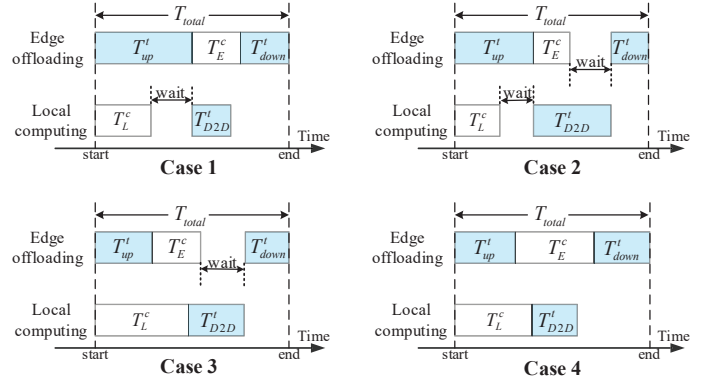}}
	\caption{The timeline of different offloading schemes.}
	\label{figure2}
	\vspace{-1em}
\end{figure}

Let us consider the process of computation and transmission more concretely. As shown in Fig.~\ref{figure2}, there are four cases in total.

{\black
\begin{itemize}
	\item \textbf{Case 1}: $T_{up}^t \geq T_L^c$  and  $T_E^c \geq T_{D2D}^t$. In this case, the local computing at user A finishes before the edge offloading. Thus user A has to wait until the task offloading is over to send the local computing result to user B through the D2D link. Moreover, in this case the transmission of the local computing result ends before the edge computing. Hence, the BS can send the edge computing results to user B directly without waiting until the transmission of D2D link is over.
	\item \textbf{Case 2}: $T_{up}^t \geq T_L^c$ and $T_E^c < T_{D2D}^t $. In this case, the local computing at user A also finishes before the edge offloading. Hence, similar with \textbf{Case 1}, user A has to wait until the task offloading is over. However, we consider that the edge computing finishes before the transmission of the local computing result. Under this situation, the BS has to wait until the D2D link transmission is over to send the edge computing results to user B. Otherwise, collisions would happen at user B.
	\item \textbf{Case 3}: $T_{up}^t < T_L^c$ and $ T_{up}^t+T_E^c < T_L^c+T_{D2D}^t $. In this case, the edge offloading ends before the local computing. Hence user A can send the local computing results to user B through the D2D link directly since the communication resource is available at the moment. Moreover, in this case the edge computing finishes before the D2D transmission of the local computing result from user A to user B. Thus, the BS has to wait until the D2D link transmission is over to send the edge computing result to user B.\footnote{It is also possible that the edge computing finishes before the local computing. However, if the BS transmits the edge computing results to user B immediately, collisions may happen because user A does not know when the BS finishes its transmission and may send the local computing results to user B simultaneously. Thus, we assume that user A has a priority to transmit results to user B compared to the BS, even if the computation at the BS ends earlier.}
	\item \textbf{Case 4}: $T_{up}^t < T_L^c$ and $ T_{up}^t+T_E^c \geq T_L^c+T_{D2D}^t$. In this case, the edge offloading ends before the local computing, and the D2D link transmission finishes before the edge computing. As a result, no wait happens.
\end{itemize}
}
According to the four cases discussed above, we obtain the expression for the overall system delay as follows,
\begin{equation}
T_{total}=
\begin{cases}
T_{up}^t+max\{T_E^c,T_{D2D}^t\}+T_{down}^t,   \\\qquad \qquad \qquad \qquad \qquad T_{up}^t\geq T_L^c,  \\
max\{T_{up}^t+T_E^c,T_{D2D}^t+T_L^c\}+T_{down}^t,   \\ \qquad \qquad \qquad \qquad \qquad T_{up}^t<T_L^c .
\end{cases}
\label{1}
\end{equation}

\subsection{Communication model}
Consider the three mmWave links that adopt hybrid A/D beamforming structures. User A and user B are equipped with $N_a$, $N_b$ antennas, respectively, \textcolor{black}{and $N_{rfa}$ $(N_{rfa}\leq N_a)$, $N_{rfb}$ $(N_{rfb}\leq N_b)$ RF chains}, respectively, while the BS has $N$ antennas and $N_{rf}$ $(N_{rf}\leq N)$ RF chains. Let $\mathbf{s}_{1} \in \mathbb{C}^{d_1\times 1}$, $\mathbf{s}_{2}\in \mathbb{C}^{d_2\times 1}$ and $\mathbf{s}_{3}\in \mathbb{C}^{d_3\times 1}$ $ \sim\mathcal{CN}(\mathbf{0},\mathbf{I})$ denote the data symbols that transmitted from user A to the BS, the BS to user B and user A to user B, respectively. The received signal at the uplink, the downlink and the D2D link can be written as\footnote{ Here we adopt a single analog beamforming matrix  $\mathbf{F}_a$ at user A and $\mathbf{F}_b$  at user B for different transmission phases, because this scheme can avoid frequent hand-off of the analog beamforming matrices with acceptable performance loss. Moreover, although we introduce the proposed algorithm under this case, it can be readily extended to the situation that there are independent analog beamforming matrices for different transmission stages.}
\begin{align*}
\mathbf{y}_{1}&=\mathbf{V}_{1}^H\mathbf{U}_{1}^H\mathbf{H}_{1}\mathbf{F} _{a}\mathbf{W}_{a1}\mathbf{s}_1+\mathbf{V}_{1}^H\mathbf{U}_{1}^H\mathbf{n}_{1}, \\
\mathbf{y}_{2}&=\mathbf{W}_{b2}^{H}\mathbf{F}_{b}^H\mathbf{H}_{2}\mathbf{V}_{2}\mathbf{U}_2\mathbf{s}_{2}+\mathbf{W}_{b2}^{H}\mathbf{F}_{b}^H\mathbf{n}_{2}, \\
\mathbf{y}_{3}&=\mathbf{W}_{b3}^{H}\mathbf{F}_{b}^H\mathbf{H}_{3}\mathbf{F}_{a}\mathbf{W}_{a3}\mathbf{s}_{3}+\mathbf{W}_{b3}^{H}\mathbf{F}_{b}^H\mathbf{n}_{3},
\end{align*}
respectively, {\black where $\mathbf{W} _{a1}\in \mathbb{C}^{N_{rfa}\times d_1}$ and $\mathbf{W} _{a3}\in \mathbb{C}^{N_{rfa}\times d_3}$ represent the transmitting digital beamforming matrices of user A for the uplink and the D2D link, respectively. $\mathbf{F} _{a}\in \mathbb{C}^{N_a\times N_{rfa}}$ represents the long-term transmitting analog beamforming matrices of user A. $\mathbf{W} _{b2} \in \mathbb{C}^{N_{rfb}\times d_2}$ and $\mathbf{W} _{b3}\in \mathbb{C}^{N_{rfb}\times d_3}$ represent the receiving digital beamforming matrices of user B for the downlink and the D2D link, respectively. $\mathbf{F} _{b}\in \mathbb{C}^{N_{b}\times N_{rfb}}$ represents the long-term receiving analog beamforming matrices of user B. $\mathbf{V}_{1}\in \mathbb{C}^{N_{rf}\times d_1}$ and $\mathbf{V}_{2}\in \mathbb{C}^{N_{rf}\times d_2}$ represent the receiving and transmitting digital beamforming vectors at the BS, respectively, and $\mathbf{U}_1 \in \mathbb{C}^{N\times N_{rf}}$ and $\mathbf{U}_2 \in \mathbb{C}^{N\times N_{rf}}$ represent the long-term receiving and transmitting analog beamforming matrices at the BS, respectively.} $\mathbf{H}_{1} \in \mathbb{C}^{N\times N_a}$, $\mathbf{H}_{2} \in \mathbb{C}^{N_b\times N}$, and $\mathbf{H}_{3} \in \mathbb{C}^{N_b\times N_a}$ denote the channel matrices of the uplink, downlink and D2D link, respectively, $\mathbb{E}\{\mathbf{n}_1\mathbf{n}_1^H\}=\sigma_1^2\mathbf{I}$, $\mathbb{E}\{\mathbf{n}_2\mathbf{n}_2^H\}=\sigma_2^2\mathbf{I}$, and $\mathbb{E}\{\mathbf{n}_3\mathbf{n}_3^H\}=\sigma_3^2\mathbf{I}$ denote the zero mean additive white Gaussian noise of the uplink, downlink and D2D link, respectively.

With the above definitions, we write the transmission rate for the uplink $R_1$, downlink $R_2$, and D2D link $R_3$, respectively as \eqref{uplinkrate}-\eqref{4}\footnote{We do not include the receiving digital beamformers in the rate expressions because it is well-known that the optimal digital receivers (i.e., minimum mean square error (MMSE) receivers) can achieve the maximum system rate, see \cite{WMMSE} for more details.},
\begin{figure*}
	\vspace{-1em}
\begin{eqnarray}
	\small
	R_{1}\!\!\!\!&=\!\!\!\!&B_{1}\log \det[\mathbf{I}+ \frac{1}{\sigma_1^2}\mathbf{U}_1^H\mathbf{H}_1\mathbf{F}_{a}\mathbf{W}_{a1}\mathbf{W}_{a1}^H\mathbf{F}_{a}^H\mathbf{H}_1^H\mathbf{U}_1(\mathbf{U}_1^H\mathbf{U}_1)^{-1}], \label{uplinkrate}\\
	R_{2}\!\!\!\!&=\!\!\!\!&B_{2}\log \det[\mathbf{I}+\frac{1}{\sigma_2^2}\mathbf{F}_{b}^H\mathbf{H}_2\mathbf{U}_2\mathbf{V}_2\mathbf{V}_2^H\mathbf{U}_2^H\mathbf{H}_2^H\mathbf{F}_{b}(\mathbf{F}_{b}^H\mathbf{F}_{b})^{-1}] \label{3},\\
	R_{3}\!\!\!\!&=\!\!\!\!&B_{3}\log \det[\mathbf{I}+\frac{1}{\sigma_3^2}\mathbf{F}_{b}^H\mathbf{H}_3\mathbf{F}_{a}\mathbf{W}_{a3}\mathbf{W}_{a3}^H\mathbf{F}_{a}^H\mathbf{H}_3^H\mathbf{F}_{b}(\mathbf{F}_{b}^H\mathbf{F}_{b})^{-1}], \label{4}
\end{eqnarray} \vspace{-1.6em}
\end{figure*}
where $B_{1}$, $B_{2}$ and $B_{3}$ represent the bandwidth of the uplink, downlink and D2D link, respectively.

{\color{black}
In practice, the relationship between the circuit power and the output power may be non-linear due to the working mode of RF power amplifiers (PA) in mmWave band~\cite{RF_PA_book}. Hence, it is necessary to take the non-linear energy efficiency of PAs into consideration. Specifically, we consider the Doherty PA in this paper, which is one of the most widely used PA architecture in high frequency band that has enhanced energy efficiency and linearity \cite{Doherty_background}. The relationship between the output power $P_{out}$ and the actual PA power consumption $P_{PA}$ is given by \cite{Doberty_expression}
\begin{equation}
	P_{PA} = \begin{cases}
		2\sqrt{P_{out}P_{max}}/\pi,  \\
		\qquad \qquad 0<P_{out}\leq 0.25P_{max}, \\
		6\sqrt{P_{out}P_{max}}/\pi-2P_{max}/\pi,  \\
		\qquad \qquad  0.25P_{max}<P_{out}<P_{max},
	\end{cases}	\label{PApower}
\end{equation}
where $P_{max}$ is the maximum output power of the PA. In order to compute the total power consumption of all PAs, we must calculate the output power of each PA first, which is given by
\begin{equation}
	P_{out1}(i) = \mathbb{E}(|\mathbf{F}_{a}(i,:)\mathbf{W}_{a1}\mathbf{s}_{1}|^2) = \|\mathbf{F}_{a}(i,:)\mathbf{W}_{a1}\|^2, \forall i \label{Pout1}
\end{equation}
\begin{equation}
	P_{out2}(i) = \mathbb{E}(|\mathbf{U}_{2}(i,:)\mathbf{V}_{2}\mathbf{s}_{2}|^2)= \|\mathbf{U}_{2}(i,:)\mathbf{V}_{2}\|^2, \forall i \label{Pout2}
\end{equation}
\begin{equation}
	P_{out3}(i) = \mathbb{E}(|\mathbf{F}_{a}(i,:)\mathbf{W}_{a3}\mathbf{s}_{3}|^2) = \|\mathbf{F}_{a}(i,:)\mathbf{W}_{a3}\|^2, \forall i \label{Pout3}
\end{equation}
where $P_{out1}(i)$ and $P_{out3}(i)$ represent the output power of the $i$th PA of user A in the uplink and D2D link, respectively, and $P_{out2}(i)$ represents the output power of the $i$th PA of the BS in the downlink. By substituting \eqref{Pout1}-\eqref{Pout3} into (\ref{PApower}), the power consumption of each PA, i.e. $P_{PA1}(i)$, $P_{PA2}(i)$ and $P_{PA3}(i)$ can be obtained.
}

\section{Problem Formulation}
\label{Section3:problem}
{\color{black}
Based on the system model introduced above, we provide the two-timescale latency minimization problem in this section. We first introduce the proposed two-timescale scheme. Then, we formulate the long-term optimization problem and the short-term optimization problem, respectively. For the former, we seek to maximize the weighted ergodic channel capacity. As for the latter, we minimize the overall system latency. The details are given as follows.

\subsection{Two-timescale scheme}
In a typical mobile radio environment, the channel
matrices appearing in the system model of Section~\ref{Section2:system} will
exhibit a random behavior and change more or less rapidly
over time. The joint design of A/D hybrid beamforming matrices and offloading ratio for each channel instance is not realistic for implementation since as it requires repeated application of a search-based algorithm with extremely high computational
complexity~\cite{tts_fd_relay}. Moreover, this approach entails a huge amount
of overhead in the estimation and exchange of real-time CSI information, and is likely to be very sensitive to CSI delays.
Therefore, to circumvent these difficulties, we hereby propose
a practical two-timescale hybrid beamforming scheme that
takes into account changes in both the instantaneous CSI and
their local statistics. Let us define the following concepts of timescales:
\begin{itemize}
	\item Long-timescale: The time interval over which the channel statistics\footnote{In this work, channel statistics refer to the moments or distribution of the channel fading realizations. The proposed long-term beamforming design only has to obtain a single (potentially outdated) channel sample at each frame. By observing one channel sample at each time, our proposed algorithm can automatically learn the channel statistics and converge to a stationary point of the considered stochastic optimization problem.} are assumed constant.
	\item Short-timescale: The time interval over which the channel gains are assumed constant, i.e. the channel coherence time.
\end{itemize}

\begin{figure*}[!t]
	\centerline{\includegraphics[width=5.2in]{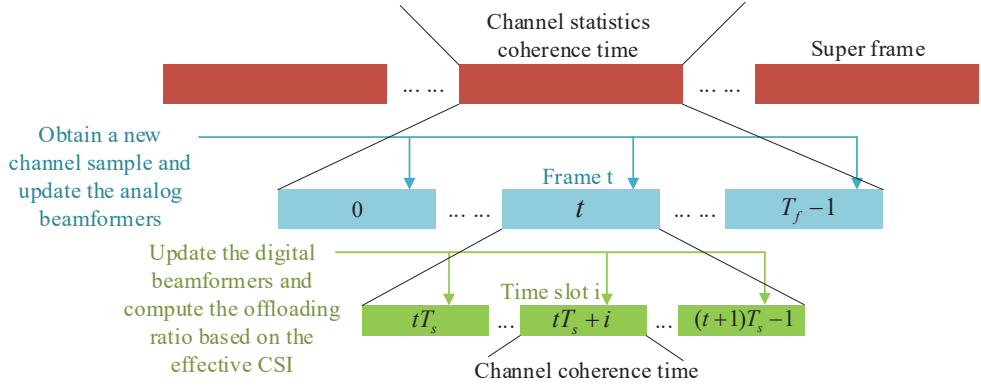}}
	\caption{The two-timescale model.}
	\label{time_scale}
	\vspace{-1em}
\end{figure*}

As illustrated in Fig.~\ref{time_scale}, the time domain is divided into a number of super frames within which the channel statistics are invariant. Each super frame consist of $T_f$ frames, and each frame is further divided into $T_s$ time slots. In our proposed
approach, to reduce CSI overhead, we only make use of one
complete estimated CSI at the end of each frame, while we
employ a so-called effective CSI (the multiplication of the analog beamforming matrices and the CSI matrices, take the uplink as an example, $\mathbf{H}_{ef1} \triangleq \mathbf{U}_1^H\mathbf{H}_1\mathbf{F}_a \in  \mathbb{C}^{N_{rf}\times N_{rfa}}$ while $\mathbf{H}_1 \in \mathbb{C}^{N \times N_a}$) with reduce dimension
within each time slot. The short-timescale digital beamforming
matrices and the offloading ratio are optimized in each time slot by using the effective
real-time channel matrices with reduced dimension, and the
long-timescale analog beamforming matrices are updated at
the end of each frame based on estimated (possibly outdated)
CSI. In the following, we formulate the long-term optimization problem
and the short-term optimization problem, respectively.
\begin{remark}
	We assume the tasks can be finished within a channel coherence time. If the user has too many tasks and cannot finish in a single time slot, then he can allocate his tasks to multiple time slots and ensure as much tasks being finished in a time slot as possible. Hence we focus on the latency minimization in a single time slot in this paper.	
\end{remark}

\vspace{-1.5em}
\subsection{Problem formulation}
Note that the long-timescale analog beamforming matrices
should be optimized based on the CSI statistics over a long-term scale, and we cannot directly optimize them by minimizing the system latency that relies on the optimal digital beamforming matrices and offloading ratio for all channel realizations. To overcome this difficulty, we propose to optimize the analog
beamforming matrices by maximizing the
weighted ergodic sum capacity, that does not depend on the short-term variables. Then, we minimize the system latency by optimizing the digital beamforming matrices and offloading ratio in each time slot.\footnote{Note that it makes sense that the maximization of the channel capacity by designing long-term analog beamforming matrices can help minimize the system latency and this formulation is more suitable for practical design. Moreover, we validate the effectiveness of the proposed algorithm in our simulation.}

1) The long-term master problem for designing analog beamforming matrices yields
\begin{equation}
	\begin{split}
    \mathcal{P} 1:\max_{\boldsymbol{\theta}_{\mathbf{U}_1},\boldsymbol{\theta}_{\mathbf{U}_2},\boldsymbol{\theta}_{\mathbf{F}_a},\boldsymbol{\theta}_{\mathbf{F}_b}} &f(\boldsymbol{\theta}_{\mathbf{U}_1},\boldsymbol{\theta}_{\mathbf{U}_2},\boldsymbol{\theta}_{\mathbf{F}_a},\boldsymbol{\theta}_{\mathbf{F}_b})\\
    &\triangleq \mathbb{E}\{g(\boldsymbol{\theta}_{\mathbf{U}_1},\boldsymbol{\theta}_{\mathbf{U}_2},\boldsymbol{\theta}_{\mathbf{F}_a},\boldsymbol{\theta}_{\mathbf{F}_b})\} \\
    &\triangleq w_1\bar{C}_1+w_2\bar{C}_2+w_3\bar{C}_3, \label{long-term objective}
\end{split}
\end{equation}
where we define $\boldsymbol{\theta}_{\mathbf{U}_1} = \angle \mathbf{U}_1$, $\boldsymbol{\theta}_{\mathbf{U}_2} = \angle \mathbf{U}_2$, $\boldsymbol{\theta}_{\mathbf{F}_a} = \angle \mathbf{F}_a$ and $\boldsymbol{\theta}_{\mathbf{F}_b} = \angle \mathbf{F}_b$ for convenience to meet the unit modulus constraint, and the weight $w_1,w_2$ and $w_3$ can be empirically chosen based on the transmission tasks of the corresponding link. Specifically, we choose the weight as $w_k = \frac{\bar{L}_k}{\sum_{i=1}^{3}\bar{L}_i}, k=1,2,3$, where $\bar{L}_1$, $\bar{L}_2$ and $\bar{L}_3$ denote the total number of transmission bits of the uplink, the downlink and the D2D link in the last super frame. Please note that this formulation is quite similar with that of maximizing the queue-length-weighted sum rate, which is widely adopted in the area of wireless resource scheduling \cite{queue_theory_1,queue_theory_2,queue_theory_3}, and it is also reasonable to use the number of transmission data in the last super frame because this information is available at the BS and the statistics between two adjacent super frames are much alike.
By defining
\begin{eqnarray}
		C_{1}\triangleq&\!\!\log \det[\mathbf{I}+ \frac{1}{\sigma_1^2}\mathbf{U}_1^H\mathbf{H}_1\mathbf{F}_{a}\mathbf{F}_{a}^H\mathbf{H}_1^H\mathbf{U}_1(\mathbf{U}_1^H\mathbf{U}_1)^{-1}], \label{uplinkcapa}\\
		C_{2}\triangleq&\!\!\!\!\log \det[\mathbf{I}+\frac{1}{\sigma_2^2}\mathbf{F}_{b}^H\mathbf{H}_2\mathbf{U}_2\mathbf{U}_2^H\mathbf{H}_2^H\mathbf{F}_{b}(\mathbf{F}_{b}^H\mathbf{F}_{b})^{-1}] \label{downlinkcapa},\\
		C_{3}\triangleq&\!\!\!\!\!\!\log \det[\mathbf{I}+\frac{1}{\sigma_3^2}\mathbf{F}_{b}^H\mathbf{H}_3\mathbf{F}_{a}\mathbf{F}_{a}^H\mathbf{H}_3^H\mathbf{F}_{b}(\mathbf{F}_{b}^H\mathbf{F}_{b})^{-1}], \label{D2Dlinkcapa}
\end{eqnarray}
then, the expressions of the ergodic channel capacity for the link between user A and the BS, the link between the BS and user B, and the link between user A and user B, are given by $\bar{C}_{1}\triangleq\mathbb{E}\{C_1\}$, $\bar{C}_{2}\triangleq\mathbb{E}\{C_2\}$, and $\bar{C}_{3}\triangleq\mathbb{E}\{C_3\}$, respectively~\cite{capacity_bound}, and  $g(\boldsymbol{\theta}_{\mathbf{U}_1},\boldsymbol{\theta}_{\mathbf{U}_2},\boldsymbol{\theta}_{\mathbf{F}_a},\boldsymbol{\theta}_{\mathbf{F}_b}) \triangleq w_1C_1+w_2C_2+w_3C_3$.

%The expressions of the ergodic channel capacity for the link between user A and the BS, the link between the BS and user B and the link between user A and user B, are given by~\cite{capacity_bound}
%\begin{eqnarray}
%	\bar{C}_{1}\triangleq&\!\!\mathbb{E}\{C_1\}\triangleq&\!\!\mathbb{E}\{\log \det[\mathbf{I}+ \frac{1}{\sigma_1^2}\mathbf{U}_1^H\mathbf{H}_1\mathbf{F}_{a}\mathbf{F}_{a}^H\mathbf{H}_1^H\mathbf{U}_1(\mathbf{U}_1^H\mathbf{U}_1)^{-1}]\}, \label{uplinkcapa}\\
%	\bar{C}_{2}\triangleq&\!\!\mathbb{E}\{C_2\}\triangleq&\!\!\mathbb{E}\{\log \det[\mathbf{I}+\frac{1}{\sigma_2^2}\mathbf{F}_{b}^H\mathbf{H}_2\mathbf{U}_2\mathbf{U}_2^H\mathbf{H}_2^H\mathbf{F}_{b}(\mathbf{F}_{b}^H\mathbf{F}_{b})^{-1}]\} \label{downlinkcapa},\\
%	\bar{C}_{3}\triangleq&\!\!\mathbb{E}\{C_3\}\triangleq&\!\!\mathbb{E}\{\log \det[\mathbf{I}+\frac{1}{\sigma_3^2}\mathbf{F}_{b}^H\mathbf{H}_3\mathbf{F}_{a}\mathbf{F}_{a}^H\mathbf{H}_3^H\mathbf{F}_{b}(\mathbf{F}_{b}^H\mathbf{F}_{b})^{-1}]\}, \label{D2Dlinkcapa}
%\end{eqnarray}
%respectively, and  $g(\boldsymbol{\theta}_{\mathbf{U}_1},\boldsymbol{\theta}_{\mathbf{U}_2},\boldsymbol{\theta}_{\mathbf{F}_a},\boldsymbol{\theta}_{\mathbf{F}_b}) \triangleq w_1C_1+w_2C_2+w_3C_3$.

2) The short-term optimization problem for designing the digital beamforming matrices and offloading ratio can be expressed as
\begin{subequations}
	\begin{align}
	\mathcal{P}2:\min_{\mathcal{S}} \,\,  & T_{total} \label{shortterm object} \\
   \text{s.t.}\,\,
    &0 \leq \rho \leq 1,  &  \label{5b}\\
    &\sum_{i=1}^{N_a}P_{PA1}(i) \leq P_{UA},\sum_{i=1}^{N_a}P_{PA3}(i) \leq P_{UA},  \label{con_UA_power}\\
    &\sum_{i=1}^{N}P_{PA2}(i) \leq P_{BS}, \label{con_BS_power} \\
    &0 \leq P_{outk}(i) \leq P_{max},  k=1,2,3,\forall i, \label{con_PA_power}
	\end{align}
\end{subequations}
where $\mathcal{S} \triangleq \{\rho,\mathbf{W}_{a1},\mathbf{W}_{a3},\mathbf{V}_2\}$ denotes the set of the short-term optimization variables. \eqref{con_UA_power} and \eqref{con_BS_power} denote the  transmit power constraints at user A and the BS, respectively. \eqref{con_PA_power} denotes the output power constraints of the PAs.

\section{Long-term analog beamforming design}
\label{Section4:longterm}
In this section, we introduce the proposed long-term analog beamforming design algorithm for solving $\mathcal{P}1$. The original problem cannot be solved straightforwardly due to the stochastic and non-convex objective function. However, based on the theoretical framework exposed in \cite{cssca}, we seek to approximate the original objective function \eqref{long-term objective} by using a quadratic surrogate function. Specifically, at the end of each channel frame $t$, the channel samples $\mathbf{H}_1^t$, $\mathbf{H}_2^t$ and $\mathbf{H}_3^t$ are obtained and the surrogate objective function is updated based on these channel samples and the approximated gradients as follows,
\begin{equation}
	\begin{split}
	\bar{f}^t(\boldsymbol{\theta}_{\mathbf{U}_1},&\boldsymbol{\theta}_{\mathbf{U}_2},\boldsymbol{\theta}_{\mathbf{F}_a},\boldsymbol{\theta}_{\mathbf{F}_b}) = \tilde{f}^t + (\mathbf{f}_{\mathbf{U}_1}^t)^T(\boldsymbol{\theta}_{\mathbf{U}_1}-\boldsymbol{\theta}_{\mathbf{U}_1}^t)\\
	&+(\mathbf{f}_{\mathbf{U}_2}^t)^T(\boldsymbol{\theta}_{\mathbf{U}_2}-\boldsymbol{\theta}_{\mathbf{U}_2}^t)+(\mathbf{f}_{\mathbf{F}_a}^t)^T(\boldsymbol{\theta}_{\mathbf{F}_a}-\boldsymbol{\theta}_{\mathbf{F}_a}^t) \\
	&+(\mathbf{f}_{\mathbf{F}_b}^t)^T(\boldsymbol{\theta}_{\mathbf{F}_b}-\boldsymbol{\theta}_{\mathbf{F}_b}^t)+\varpi\|\boldsymbol{\theta}_{\mathbf{U}_1}-\boldsymbol{\theta}_{\mathbf{U}_1}^t\|^2\\
	&+\varpi\|\boldsymbol{\theta}_{\mathbf{U}_2}-\boldsymbol{\theta}_{\mathbf{U}_2}^t\|^2
	+\varpi\|\boldsymbol{\theta}_{\mathbf{F}_a}-\boldsymbol{\theta}_{\mathbf{F}_a}^t\|^2\\
	&+\varpi\|\boldsymbol{\theta}_{\mathbf{F}_b}-\boldsymbol{\theta}_{\mathbf{F}_b}^t\|^2,	\label{surrogate function}
    \end{split}
\end{equation}
where $\varpi$ is a constant, and $\tilde{f}^t$, $\mathbf{f}_{\mathbf{U}_1}^t$, $\mathbf{f}_{\mathbf{U}_2}^t$, $\mathbf{f}_{\mathbf{F}_a}^t$ and $\mathbf{f}_{\mathbf{F}_b}$ denote the approximation of the objective function $f$, the partial derivatives $\frac{\partial f}{\partial \boldsymbol{\theta}_{\mathbf{U}_1}}$, $\frac{\partial f}{\partial \boldsymbol{\theta}_{\mathbf{U}_2}}$, $\frac{\partial f}{\partial \boldsymbol{\theta}_{\mathbf{F}_a}}$ and $\frac{\partial f}{\partial \boldsymbol{\theta}_{\mathbf{F}_b}}$, respectively. The quantities can be updated based on the following expressions
\begin{equation}
	\tilde{f}^t = (1-\varepsilon^t)\tilde{f}^{t-1} - \varepsilon^tg(\boldsymbol{\theta}_{\mathbf{U}_1}^t,\boldsymbol{\theta}_{\mathbf{U}_2}^t,\boldsymbol{\theta}_{\mathbf{F}_a}^t,\boldsymbol{\theta}_{\mathbf{F}_b}^t), \label{surrogate_f}
\end{equation}
\begin{equation}
	\mathbf{f}_{\mathbf{U}_1}^t = (1-\varepsilon^t)\mathbf{f}_{\mathbf{U}_1}^{t-1} - \varepsilon^t \frac{\partial g}{\partial \boldsymbol{\theta}_{\mathbf{U}_1} }(\boldsymbol{\theta}_{\mathbf{U}_1}^t,\boldsymbol{\theta}_{\mathbf{U}_2}^t,\boldsymbol{\theta}_{\mathbf{F}_a}^t,\boldsymbol{\theta}_{\mathbf{F}_b}^t),
\end{equation}
\begin{equation}
	\mathbf{f}_{\mathbf{U}_2}^t = (1-\varepsilon^t)\mathbf{f}_{\mathbf{U}_2}^{t-1} - \varepsilon^t \frac{\partial g}{\partial \boldsymbol{\theta}_{\mathbf{U}_2} }(\boldsymbol{\theta}_{\mathbf{U}_1}^t,\boldsymbol{\theta}_{\mathbf{U}_2}^t,\boldsymbol{\theta}_{\mathbf{F}_a}^t,\boldsymbol{\theta}_{\mathbf{F}_b}^t),
\end{equation}
\begin{equation}
\mathbf{f}_{\mathbf{F}_a}^t = (1-\varepsilon^t)\mathbf{f}_{\mathbf{F}_a}^{t-1} - \varepsilon^t \frac{\partial g}{\partial \boldsymbol{\theta}_{\mathbf{F}_a} }(\boldsymbol{\theta}_{\mathbf{U}_1}^t,\boldsymbol{\theta}_{\mathbf{U}_2}^t,\boldsymbol{\theta}_{\mathbf{F}_a}^t,\boldsymbol{\theta}_{\mathbf{F}_b}^t),
\end{equation}
\begin{equation}
	\mathbf{f}_{\mathbf{F}_b}^t = (1-\varepsilon^t)\mathbf{f}_{\mathbf{F}_b}^{t-1} - \varepsilon^t \frac{\partial g}{\partial \boldsymbol{\theta}_{\mathbf{F}_b} }(\boldsymbol{\theta}_{\mathbf{U}_1}^t,\boldsymbol{\theta}_{\mathbf{U}_2}^t,\boldsymbol{\theta}_{\mathbf{F}_a}^t,\boldsymbol{\theta}_{\mathbf{F}_b}^t), \label{surrogate_Fb}
\end{equation}
with initial value $\tilde{f}^{-1} = 0$, $\mathbf{f}_{\mathbf{U}_1}^{-1} = \mathbf{0}$, $\mathbf{f}_{\mathbf{U}_2}^{-1} = \mathbf{0}$, $\mathbf{f}_{\mathbf{F}_a}^{-1} = \mathbf{0}$ and $\mathbf{f}_{\mathbf{F}_b}^{-1} = \mathbf{0}$. The expressions of the partial derivatives are given in \textbf{Appendix A}, and $\{\varepsilon^t\}$ is a sequence of the parameters to be properly chosen.
Subsequently, we aim to solve the approximated problem at time frame $t$, which is given by
\begin{equation}
	\min_{\boldsymbol{\theta}_{\mathbf{U}_1},\boldsymbol{\theta}_{\mathbf{U}_2},\boldsymbol{\theta}_{\mathbf{F}_a},\boldsymbol{\theta}_{\mathbf{F}_b}} \bar{f}^t(\boldsymbol{\theta}_{\mathbf{U}_1},\boldsymbol{\theta}_{\mathbf{U}_2},\boldsymbol{\theta}_{\mathbf{F}_a},\boldsymbol{\theta}_{\mathbf{F}_b}).
\end{equation}
It is readily seen that this is a convex problem and the solution is given by
\begin{equation}
	\begin{split}
	&\bar{\boldsymbol{\theta}}_{\mathbf{U}_1} = \boldsymbol{\theta}_{\mathbf{U}_1}^t-\frac{\mathbf{f}_{\mathbf{U}_1}^t}{2\varpi}, \bar{\boldsymbol{\theta}}_{\mathbf{U}_2}= \boldsymbol{\theta}_{\mathbf{U}_2}^t-\frac{\mathbf{f}_{\mathbf{U}_2}^t}{2\varpi}, \\
	&\bar{\boldsymbol{\theta}}_{\mathbf{F}_a} = \boldsymbol{\theta}_{\mathbf{F}_a}^t-\frac{\mathbf{f}_{\mathbf{F}_a}^t}{2\varpi},
	\bar{\boldsymbol{\theta}}_{\mathbf{F}_b} = \boldsymbol{\theta}_{\mathbf{F}_b}^t-\frac{\mathbf{f}_{\mathbf{F}_b}^t}{2\varpi}. \label{solution_to_surrogate}
\end{split}
\end{equation}
Then, the long-term variables are updated as
\begin{equation}
	\begin{split}
	&\boldsymbol{\theta}_{\mathbf{U}_1}^{t+1} = (1-\gamma^t)\boldsymbol{\theta}_{\mathbf{U}_1}^{t} + \gamma^t\bar{\boldsymbol{\theta}}_{\mathbf{U}_1},
	\boldsymbol{\theta}_{\mathbf{U}_2}^{t+1} = (1-\gamma^t)\boldsymbol{\theta}_{\mathbf{U}_2}^{t} + \gamma^t\bar{\boldsymbol{\theta}}_{\mathbf{U}_2}, \\
	&\boldsymbol{\theta}_{\mathbf{F}_a}^{t+1} = (1-\gamma^t)\boldsymbol{\theta}_{\mathbf{F}_a}^{t} + \gamma^t\bar{\boldsymbol{\theta}}_{\mathbf{F}_a},
	\boldsymbol{\theta}_{\mathbf{F}_b}^{t+1} = (1-\gamma^t)\boldsymbol{\theta}_{\mathbf{F}_b}^{t} + \gamma^t\bar{\boldsymbol{\theta}}_{\mathbf{F}_b}, 	\label{longterm_update}
     \end{split}
\end{equation}
where similarly $\{\gamma_t\}$ denotes a sequence of parameters. Based on \cite{cssca}, the convergence can be guaranteed if we choose $\varepsilon^t$ and $\gamma^t$ by following the conditions
\begin{equation}
	\begin{split}
	&\lim_{t\longrightarrow \infty} \varepsilon^t = 0, \sum_{t} \varepsilon^t  = \infty, \sum_{t} (\varepsilon^t)^2  < \infty,\\
	&\lim_{t\longrightarrow \infty} \gamma^t = 0, \sum_{t} \gamma^t  = \infty, \sum_{t} (\gamma^t)^2  < \infty, \lim_{t\longrightarrow \infty} \frac{\gamma^t}{\varepsilon^t} = 0.
	\end{split}
\end{equation}
Then the proposed SSCA-based algorithm can be guaranteed to converge to a stationary solution of $\mathcal{P}1$. We summarize the proposed long-term analog beamforming design algorithm in \textbf{Algorithm 1}, and its complexity is dominated by the procedure of updating the surrogate functions, which is given by $\mathcal{O}\{N_{rf}^3+NN_{rf}(N_a+N_b)\}$.

\begin{algorithm}[t]\caption{Proposed SSCA-based algorithm for the long-term analog beamforming design}
	\begin{algorithmic}[1]
		\footnotesize
		\begin{small}
			\STATE Initialize the optimization variables $\boldsymbol{\theta}_{\mathbf{U}_1}^0,\boldsymbol{\theta}_{\mathbf{U}_2}^0,\boldsymbol{\theta}_{\mathbf{F}_a}^0,\boldsymbol{\theta}_{\mathbf{F}_b}^0$ with a feasible point. Set an appropriate value for $\varpi$ and let $t=0$.
			\REPEAT
			\STATE Obtain the CSI samples $\mathbf{H}_1^t$, $\mathbf{H}_2^t$ and $\mathbf{H}_3^t$. Compute the surrogate function \eqref{surrogate function} based on \eqref{surrogate_f}$\text{-}$\eqref{surrogate_Fb} and $\varepsilon^t$.
			\STATE Obtain the optimal solution via \eqref{solution_to_surrogate}.
			\STATE Update $\boldsymbol{\theta}_{\mathbf{U}_1}^t,\boldsymbol{\theta}_{\mathbf{U}_2}^t,\boldsymbol{\theta}_{\mathbf{F}_a}^t,\boldsymbol{\theta}_{\mathbf{F}_b}^t$ based on \eqref{longterm_update} and $\gamma^t$.
			\STATE Update the iteration number $t=t+1$.
			\UNTIL the convergence condition is satisfied or the maximum number of iterations is reached.
		\end{small}
	\end{algorithmic}
\end{algorithm}
}

\section{short-term digital beamforming and offloading ratio design}
\label{Section5:shortterm}
In this section, we introduce the proposed algorithm for solving $\mathcal{P}$2. We first decompose $\mathcal{P}$2 into several subproblems that are easier to solve. Then, for the subproblems regarding the digital beamforming design, we propose a penalty-CCCP based algorithm to handle the non-linear power consumption constraints. As for the subproblems regarding the offloading ratio design, we derive closed-form solution via classification and discussion. The details are as follows.

\subsection{Problem decomposition}\label{AA}
Due to the fact that the transmissions occur over orthogonal time in a single time slot, the transmission rates of the uplink, downlink and D2D link, i.e. $R_{1}$, $R_{2}$, $R_{3}$ are independent of each other. Furthermore, since the overall system delay (\ref{shortterm object}) is nonincreasing with $R_{1}$, $R_{2}$ and $R_{3}$, we can maximize the transmission rates first, and then optimize the offloading ratio. Hence the short-term latency minimization problem $\mathcal{P}$2 can be equivalently decomposed into the following two parts.

\begin{itemize}
	\item The digital beamforming design subproblems for the uplink, the downlink and the D2D link, respectively, are provided as: \vspace{-1mm}
	\begin{subequations}
		\begin{align}
			\mathcal{P}3\text{-}1: \max\limits_{\left\{\mathbf{W}_{a1}\right\}} &\quad R_{1}\\
			\mbox{s.t.} \quad &\sum_{i=1}^{N_a}P_{PA1}(i) \leq P_{UA}, \label{uplink_PA_constraint}\\
			&0 \leq \|\mathbf{F}_a(i,:)\mathbf{W}_{a1}\|^2 \leq P_{max}, \forall i, \label{uplink_PA_maxpower}
		\end{align}	
	\end{subequations}
	\begin{subequations} 	
	\begin{align}
	\mathcal{P} 3\text{-}2: \max\limits_{\left\{\mathbf{V}_{2}\right\}} &\quad R_{2} \\
	\mbox{s.t.} \quad &\sum_{i=1}^{N}P_{PA2}(i) \leq P_{BS},\\
	&0 \leq \|\mathbf{U}_2(i,:)\mathbf{V}_{2}\|^2 \leq P_{max},\forall i,
	\end{align}		
    \end{subequations}
	\begin{subequations}		
	\begin{align}
	\mathcal{P} 3\text{-}3: \max\limits_{\left\{\mathbf{W}_{a3}\right\}} &\quad R_{3} \\
	\mbox{s.t.} \quad &\sum_{i=1}^{N_a}P_{PA3}(i) \leq P_{UA} ,\\
	&0 \leq \|\mathbf{F}_a(i,:)\mathbf{W}_{a3}\|^2 \leq P_{max},\forall i,
	\end{align}	
	\end{subequations}
	\item The offloading ratio optimization subproblem is given by: \vspace{-0.5em}
	\begin{equation}
	\mathcal{P} 4: \min_{0 \leq \rho \leq 1} \,\, T_{total}.  \label{9} \vspace{-0.5em}
	\end{equation}
\end{itemize}

Although we have decomposed the original problem into several more tractable one, there are still some challenges, i.e. the non-linear power consumption constraint in $\mathcal{P}3$ and the multi-case piece-wise objective function in $\mathcal{P}4$. In the following two subsections, we introduce our proposed algorithms for tackling these issues.
\vspace{-1mm}
\subsection{Short-term digital beamforming design}
{\color{black}
In this subsection, we introduce the proposed short-term digital beamformer design algorithm for solving $\mathcal{P} 3\text{-}1$$ - $$\mathcal{P} 3\text{-}3$. Since these subproblems are essentially the same, we focus on the uplink to
introduce our proposed algorithm. First, we equivalently transform $\mathcal{P} 3\text{-}1$ into a more tractable form based on the celebrated weighted minimum mean square error (WMMSE) method \cite{WMMSE} as follows,
\begin{subequations}
	\label{wmmse_problem}
	\begin{align}
		\min_{\mathbf{V}_1,\mathbf{W_{a1}},\mathbf{Z}} \quad & \text{Tr}(\mathbf{Z}\mathbf{E}) - \log \det(\mathbf{Z})   \\
		\text{s.t.}\quad& \eqref{uplink_PA_constraint},\eqref{uplink_PA_maxpower},
	\end{align}
\end{subequations}
where $\mathbf{Z}$ is an auxiliary variable satisfying $\mathbf{Z}\succeq\mathbf{0}$ and
\begin{equation}
	\begin{split}	
	\mathbf{E} \triangleq& (\mathbf{V}_1^H\mathbf{H}_{ef1}\mathbf{W}_{a1} - \mathbf{I})(\mathbf{V}_1^H\mathbf{H}_{ef1}\mathbf{W}_{a1} - \mathbf{I})^H \\&+\sigma_1^2\mathbf{V}_1^H\mathbf{U}_1^H\mathbf{U}_1\mathbf{V}_1. \label{E_expression}
\end{split}
\end{equation}
Then, to tackle the non-linear power constraint, we introduce two auxiliary variables $P_{PA1}$ and $V_{out1}$, and equivalently convert \eqref{wmmse_problem} as
\begin{subequations}
	\begin{align}
		\min_{\bar{\mathcal{S}}} \quad & \text{Tr}(\mathbf{Z}\mathbf{E}) - \log \det(\mathbf{Z})  \label{P5_obj} \\
		\text{s.t.}\quad& \sum_{i=1}^{N_a}P_{PA1}(i) = \bar{P}_{UA}, \label{con_total_power}  \\
		&0<P_{PA1}(i)\leq 4 P_{max}/\pi, \forall i, \label{con_Ppai}\\
		&0<V_{out1}(i) \leq P_{max}, \forall i, \label{con_Vouti}\\
		&P_{PA1}(i) = h(V_{out1}(i)), \forall i,\label{con_eq_f}\\
		&V_{out1}(i) = \|\mathbf{F}_{a}(i,:)\mathbf{W}_{a1}\|, \forall i, \label{con_eq_Vouti}
	\end{align}
\end{subequations}
where $\bar{\mathcal{S}} \triangleq \{\mathbf{V}_1,\mathbf{W_{a1}},\mathbf{Z},P_{PA1},V_{out1}\}$ is the set of optimization variables and $\bar{P}_{UA} \triangleq \min(P_{UA}$,$4N_aP_{max}/\pi)$, while $h(V_{out1}(i))$ is defined as
\begin{equation}
	h(V_{out1}(i)) = \begin{cases}
		2V_{out1}(i)\sqrt{P_{max}}/\pi, \\
		\qquad \qquad  0<V_{out1}(i)\leq 0.25P_{max}, \\
		6V_{out1}(i)\sqrt{P_{max}}/\pi-2P_{max}/\pi, \\
		\qquad \qquad 0.25P_{max}<V_{out1}(i)<P_{max}.
	\end{cases}	\label{func_f}
\end{equation}
Then, we solve the transformed problem based on the penalty-CCCP framework, the detailed introduction of which can be found in \cite{pcccp}. By penalizing the equality constraints \eqref{con_total_power}, \eqref{con_eq_f} and \eqref{con_eq_Vouti} into the objective function \eqref{P5_obj}, we obtain the penalized problem shown as follows.
\begin{subequations}
	\begin{align}
		\mathcal{P}5:\min_{\mathcal{S}} &\quad  \text{Tr}(\mathbf{Z}\mathbf{E}) - \log \det(\mathbf{Z}) \nonumber\\
		&+\frac{1}{2\varrho}(\sum_{i=1}^{N_a}(P_{PA1}(i)-h(V_{out1}(i)))^2\nonumber  \\ &+\sum_{i=1}^{N_a}(\|\mathbf{F}_a(i,:)\mathbf{W}_{a1}\|-V_{out1}(i))^2\nonumber\\
		&+(\sum_{i=1}^{N_a}P_{PA1}(i)-\bar{P}_{UA})^2) \label{P6_obj} \\
		\text{s.t.} &\quad  \eqref{con_Ppai},\eqref{con_Vouti}, \nonumber
	\end{align}
\end{subequations}
where $\varrho$ denotes a penalty coefficient. Referring to the penalty-CCCP, the proposed algorithm contains two loops, where the penalty coefficient is adjusted in the outer loop, while in the inner loop the optimization variables are updated in a block coordinate descent fashion. In each block of the inner loop, we aim to decompose the resulting penalized problem into a number of subproblems, which can be solved easily in parallel. To this end, we divide the optimization variables into three blocks, i.e. $\{\mathbf{V}_1,P_{PA1}\}$, $\{\mathbf{Z},V_{out1}\}$, $\{\mathbf{W}_{a1}\}$. The detailed solutions within each block are given in \textbf{Appendix B} and we summarize the proposed penalty-CCCP based algorithm for uplink short-term digital beamforming design in \textbf{Algorithm 2}. The complexity is given by $\mathcal{O}\{I_1I_2(N_{rf}^3+N_{rfa}^3+NN_{rf}N_a)\}$, where $I_1$ and $I_2$ denote the iteration numbers for the outer and inner loops, respectively.

Although the computational complexity in a single iteration is low, the required iteration number may be large for the double loop nature of the penalty-CCCP. Hence, we also propose a low-complexity heuristic algorithm for the short-term digital beamformer design. Ignoring the non-linear power constraint, the uplink digital beamforming design problem yields
\begin{subequations}
	\begin{align}
	\max_{\mathbf{W}_{a1}} \quad & \log \det(\mathbf{I}+\frac{1}{\sigma_1^2}\mathbf{H}_{ef1}\mathbf{W}_{a1}\mathbf{W}_{a1}^H\mathbf{H}_{ef1}^H(\mathbf{U}_1^H\mathbf{U}_1)^{-1}) \\
	\text{s.t.} \quad & \text{Tr}\{\mathbf{Q}_{Fa}\mathbf{W}_{a1}\mathbf{W}_{a1}^H\} \leq P_{UA},
	\end{align}
\end{subequations}
where $\mathbf{Q}_{Fa} \triangleq \mathbf{F}_a^H\mathbf{F}_a$. Since the long-term analog beamformer $\mathbf{U}_1$ is fixed, we further view $\bar{\mathbf{H}}_{ef1} \triangleq \bar{\mathbf{\Sigma}}^{-1}\bar{\mathbf{V}}\mathbf{H}_{ef1}$ as the effective channel matrix, where $\bar{\mathbf{\Sigma}}$ and $\bar{\mathbf{V}}$ are the diagonal singular value matrix and the right singular vector matrix  of $\mathbf{U}_1$, respectively. This problem has a well-known water-filling solution which is given by
\begin{equation}
	\mathbf{W}_{a1} \triangleq \mathbf{Q}_{Fa}^{-1/2}\mathbf{U}_e\mathbf{\Sigma}_e,
\end{equation}
where $\mathbf{U}_e$ is the set of the right singular vectors corresponding to the $N_a$ largest singular values of $\bar{\mathbf{H}}_{ef1}\mathbf{Q}_{Fa}^{-1/2}$, and $\mathbf{\Sigma}_e$ is the diagonal power allocation matrix. Since this low-complexity algorithm does not consider the nonlinear transmit power constraint directly, we scale the digital beamforming matrix $\mathbf{W}_{a1}$ to satisfy constraint \eqref{uplink_PA_constraint} and \eqref{uplink_PA_maxpower}, where the scaling factor can be conveniently found by using the bisection search. The computational complexity of the proposed low-complexity short-term digital beamforming design algorithm is given by $\mathcal{O}\{NN_aN_{rf}+N_{rf}N_{rfa}^2\}$.

\begin{algorithm}[t]\caption{Proposed penalty-CCCP based algorithm for uplink short-term digital beamformer design}
	\begin{algorithmic}[1]
		\footnotesize
		\begin{small}
			\STATE Initialize the optimization variables with a feasible point. Define the tolerance of accuracy $\delta_1$ and $\delta_2$. Set iteration number $i=0$ and $j=0$. Set  $\varrho^{0} >0$ and $c>1$.
			\REPEAT
			\REPEAT
			\STATE update $\mathbf{V}_1$ and $P_{PA1}$ according to (\ref{optimal_V1}) and (\ref{optimal_PPA}),  respectively.
			\STATE update $\mathbf{Z}$ and $V_{out1}$ according to (\ref{optimalZ}) and (\ref{optimal_Vout}), respectively.
			\STATE update $\mathbf{W}_{a1}$ according to (\ref{optimal_W_a}).
			\STATE update the iteration number $i=i+1$.
			\UNTIL the difference between two successive objective values is less than $\delta_1$.
			\STATE $\varrho^{j+1}=c\varrho^{j}$.
			\STATE update the iteration number $j=j+1$.
			\UNTIL the difference of successive objective function value is less than $\delta_1$ and the penalty term is less than $\delta_2$ or the maximum number of iterations is reached.
		\end{small}
	\end{algorithmic}
\end{algorithm}
}
\vspace{-1mm}
\subsection{Optimization of offloading ratio $\rho$}
\label{Section5:offloading}
In this subsection, we aim to optimize the offloading ratio by solving $\mathcal{P}$4. Referring to the analysis of different cases for timelines shown in Fig. 2, it is readily seen that $T_{total}$ is a linear piece-wise function of $\rho$, and we can derive the optimal expression for $\rho$ through classification and analysis. Based on the four cases shown in Fig. \ref{figure2}, let us rewrite the expression of $T_{total}$ with $\rho$ being the variable as (\ref{Ttotal_4case}). In order to derive the optimal $\rho$, we need to analyze all possible situations. It is apparent that when $\rho$ grows from 0 to 1, \textbf{Case 1} and \textbf{Case 3} will happen for sure, while the occurrence of \textbf{Case 2} depends on the condition of $\frac{K_L}{K_{1}+K_L} < \frac{K_{3}}{K_{3}+K_E}$, which can be simplified to $\frac{K_L}{K_{1}} < \frac{K_{3}}{K_E}$. Similar to \textbf{Case 2}, the occurrence of \textbf{Case 4} depends on $\frac{K_L}{K_{1}+K_L} \geq \frac{K_{3}+K_L}{K_{3}+K_L+K_{1}+K_E}$, which can be simplified to $\frac{K_L}{K_{1}} \geq \frac{K_{3}}{K_E} $. As we can see, the criteria of \textbf{Case 2} and \textbf{Case 4} contradict each other and thus only one of them can appear. Hence, there are two possible situations in general.
	
\begin{figure*}[!t]
	\begin{subequations}
		\begin{numcases}{T_{total}=}
			\!\!\textbf{Case 1}:\: K_{1}\rho +K_E\rho+K_{2}\rho, &$\text{if} \,\, \rho \in [\frac{K_L}{K_{1}+K_L},1] \cap    [\frac{K_{3}}{K_{3}+K_E},1],$ \\
			\!\!\text{\textbf{Case 2}}: \: K_{1}\rho +K_{3}(1-\rho)+K_{2}\rho, & $\text{if}  \,\, \rho \in [\frac{K_L}{K_{1}+K_L},1] \cap    [0,\frac{K_{3}}{K_{3}+K_E}),$ \\
			\!\!\text{\textbf{Case 3}}: \:  (K_L +K_{3})(1-\rho)+K_{2}\rho, & $\text{if}  \,\,  \rho \in [0,\frac{K_L}{K_{1}+K_L}) \cap    [0,\frac{K_{3}+K_L}{K_{3}+K_L+K_{1}+K_E}),\qquad\qquad\quad$ \\
			\!\!\text{\textbf{Case 4}}: \: K_{1}\rho +K_E\rho+K_{2}\rho, &$\text{if}  \,\,   \rho \in [0,\frac{K_L}{K_{1}+K_L}) \cap [\frac{K_{3}+K_L}{K_{3}+K_L+K_{1}+K_E},1].$
		\end{numcases}
		\label{Ttotal_4case}
	\end{subequations}
\end{figure*}

i)\textit{ Situation A}: $\frac{K_L}{K_{1}} \geq \frac{K_{3}}{K_E} $

In this situation, \textbf{Case 2} does not happen. Thus, the expression of $T_{total}$ consists of (\ref{Ttotal_4case}c), (\ref{Ttotal_4case}d) and (\ref{Ttotal_4case}a) in order as $\rho$ increases from 0 to 1. However, note that the line function (\ref{Ttotal_4case}a) is the same as (\ref{Ttotal_4case}d). Therefore, the expression of $T_{total}$ in situation A essentially consists of two line segments. It is apparent that (\ref{Ttotal_4case}a) or (\ref{Ttotal_4case}d) is nondecreasing, while the monotonicity of (\ref{Ttotal_4case}c) depends on whether $K_{2}-K_L-K_{3}$ is positive or negative:
\begin{itemize}
	\item[1)] $K_{2}-K_L-K_{3} \geq 0$: In this case, (\ref{Ttotal_4case}c) is a nondecreasing function of $\rho$, thus the whole function is nondecreasing and we have $\rho^*=0$.
	\item[2)] $K_{2}-K_L-K_{3} < 0$: In this case, the expression of $T_{total}$  consists of a decreasing line followed by an increasing one, thus the optimal $\rho$ should be at the turning point, i.e. $\rho^*=\frac{K_{3}+K_L}{K_{3}+K_L+K_{1}+K_E} $.
\end{itemize}

ii)\textit{ Situation B}: $\frac{K_L}{K_{1}} < \frac{K_{3}}{K_E} $

In this situation, \textbf{Case 4} does not happen. So the expression of $T_{total}$ consists of (\ref{Ttotal_4case}c), (\ref{Ttotal_4case}b) and (\ref{Ttotal_4case}a) in order. Since (\ref{Ttotal_4case}a) is a nondecreasing line segment, we can conclude that $\rho^* \in [0,\frac{K_{3}}{K_{3}+K_E}]$. Then, let us focus on the monotonicity of (\ref{Ttotal_4case}c) and (\ref{Ttotal_4case}b). There are four kinds of situations in total:

\begin{itemize}
	\item[1)]$K_{2}-K_L-K_{3} < 0$ and $K_{2}+K_{1}-K_{3} < 0$:
	In this case, $T_{total}$ is monotonically decreasing when $\rho \in [0,\frac{K_{3}}{K_{3}+K_E}]$. Thus we obtain $\rho^*=\frac{K_{3}}{K_{3}+K_E}$.
	\item[2)]$K_{2}-K_L-K_{3} < 0$ and $K_{2}+K_{1}-K_{3} \geq 0$:
	In this case, $T_{total}$ is decreasing first when $\rho \in [0,\frac{K_{L}}{K_{1}+K_L}]$ and then increasing when $\rho \in [\frac{K_{L}}{K_{1}+K_L},\frac{K_{3}}{K_{3}+K_E}]$. Thus we obtain $\rho^*=\frac{K_{L}}{K_{1}+K_L}$.
	\item[3)]$K_{2}-K_L-K_{3} \geq 0$ and $K_{2}+K_{1}-K_{3} < 0$:
	This case is impossible because from $K_{2}-K_L-K_{3} \geq 0$ we obtain $K_{2} \geq K_{3}$ while from $K_{2}+K_{1}-K_{3} < 0$ we obtain $K_{2} < K_{3}$.
	\item[4)]$K_{2}-K_L-K_{3} \geq 0$ and $K_{2}+K_{1}-K_{3} \geq 0$:
	In this case, $T_{total}$ is nondecreasing in the whole feasible region of $\rho$. Thus we obtain $\rho^*=0$.
\end{itemize}

\begin{figure}[t]
	\centerline{\includegraphics[width=3.5in]{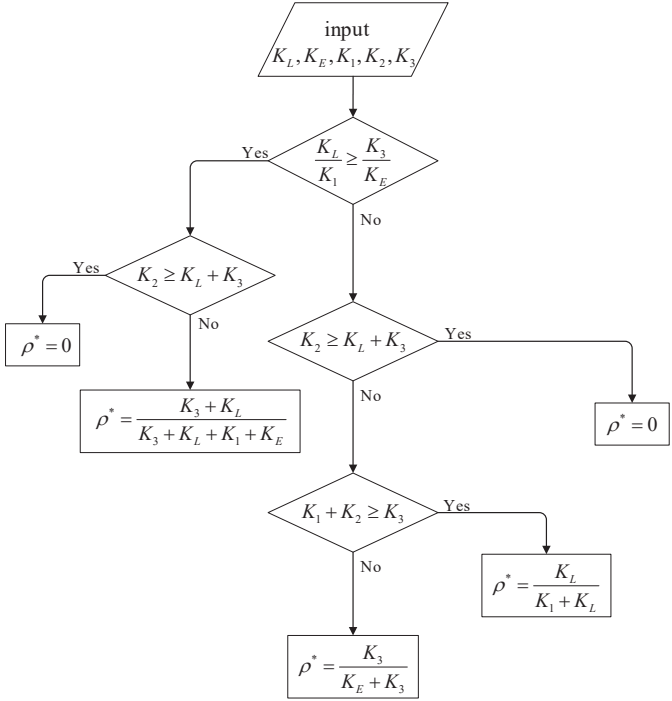}}
	\caption{Offloading ratio optimization. }
	\label{rhoexpression} \vspace{-1.5em}
\end{figure}

By following the above discussion, we obtain the final result of the optimal $\rho$. The above analysis is summarized in a flowchart given as Fig. \ref{rhoexpression}.

The complexity of the proposed short-term variables design algorithm for solving $\mathcal{P}2$ is dominated by the penalty-CCCP algorithm, whose complexity is given above. Regarding the convergence, according to the detailed convergence analysis of the penalty-CCCP algorithm~\cite{pcccp}, the proposed short-term digital beamforming algorithms converge to the stationary solutions of $\mathcal{P}3\text{-}1$, $\mathcal{P}3\text{-}2$ and $\mathcal{P}3\text{-}3$. Moreover, considering the optimality of the derived offloading ratio and the fact that $T_{total}$ is non-increasing with the transmission rate $R_1$, $R_2$ and $R_3$. The proposed joint short-term digital beamforming and offloading ratio design algorithm converge to a stationary point of $\mathcal{P}2$.

\begin{remark}
	We assume that the design algorithm is implemented at the BS. Specifically, in each time slot, the effective uplink CSI matrix $\mathbf{H}_{ef1}$ is estimated at the BS. The effective downlink CSI matrix $\mathbf{H}_{ef2}$ and D2D CSI matrix $\mathbf{H}_{ef3}$ are estimated at user B and fed back to the BS. Moreover, user A sends the necessary information such as the local computing capacity $F_{L}$ and the task compression ratio $\alpha$ to the BS and the BS conducts the short-term digital beamforming and task allocation algorithm. In each frame, the full channel samples $\mathbf{H}_1$, $\mathbf{H}_2$ and $\mathbf{H}_3$ are collected by the BS using the similar channel estimation strategy and \textbf{Algorithm 1} is then performed.
\end{remark}

\section{Simulation results}
\label{Section6:simulation}
In this section, we present simulation results to verify the effectiveness of our proposed algorithm. The simulation parameters are provided as follows unless otherwise stated. \textcolor{black}{The number of antennas at the BS server is set as $N=64$, while the number of antennas at the users is $N_a=N_b=8$. The number of RF chains at the BS is $N_{rf}=4$, with  user RF chain numbers set as $N_{rfa}=N_{rfb}=2$. We set the number of data streams as $d_1 = \min(N_{rfa},N_{rf})$, $d_2 = \min(N_{rf},N_{rfb})$ and $d_3 = \min(N_{rfa},N_{rfb})$, respectively. For the mmWave channel model, we employ the generally used extended Salch-Valenzuela geometric model \cite{mmWave_channelmodel}. Specifically, the channel matrix is given by
\begin{equation}
\mathbf{H}=\sqrt{\frac{N_1N_2}{L_p}}\sum_{l=1}^{L_p}\alpha_l\textbf{\text{a}}_r(\varphi_l^r)\textbf{\text{a}}_t(\varphi_l^t)^H \times \exp(j2\pi f_d\tau\cos(\varphi_l^r)),
\end{equation}
where $N_1$ and $N_2$ are the number of transmit and receive antennas, respectively. $L_p$ is the number of distinguishable paths, $\alpha_l \sim \mathcal{CN}(0,\sigma_{pl}^2)$ is the complex gain of the $l\text{-}$th path, $\textbf{\text{a}}_r(\varphi_l^r)$ and $\textbf{\text{a}}_t(\varphi_l^t)$ are the receive and transmit antenna array response vectors, where $\varphi_l^r$ and $\varphi_l^t$ are the azimuth angles of arrival and departure, respectively. $f_d$ is the maximum Doppler shift, and $\tau$ is the delay. The expression of the response vector is given by \begin{equation}
	\textbf{\text{a}}(\theta)=\frac{1}{N}\left[1,e^{jk_0d_a\pi\sin(\theta)},...,e^{jk_0d_a(N-1)\pi\sin(\theta)}\right]^T,
\end{equation}
where $k_0 = 2\pi/\lambda_0$, $\lambda_0$ is the wavelength and $d_a$ is the antenna spacing set as $d_a = \lambda_0/2$. We assume that there are 1 line-of-sight (LOS) path and 15 non-line-of-sight (NLOS) path, and the gain for the LOS path is $\sigma_{p1}^2 = 1$, while the gain for the NLOS path is $\sigma_{pl}^2 = 0.1, \forall l\neq 1$. We set the Doppler shift as $f_d = 70\text{Hz}$ and the transmission delay as $\tau = 4\text{ms}$ according to \cite{tts_fd_relay}.}

{\color{black}The BS is located at $[0,0,10\text{m}]$ and the positions of user A and user B are set to $[D_x,D_y,1\text{m}]$ and $[-D_x,D_y,1\text{m}]$, respectively, with $D_x = 5\text{m}$ and $D_y = 50\text{m}$. The path loss is modeled as $P_{ls} = C_0(\frac{d_{link}}{D_0})^{-\beta}$, where $C_0$ is the path loss at the reference distance $D_0 = 1\text{m}$ and is set to $C_0 = -30$dB, $d_{link}$ is the link distance, and $\beta$ is the path loss exponent where we set it for the uplink, the downlink and the D2D link as $\beta_1 = 3$, $\beta_2 = 3$ and $\beta_3 = 2.4$, respectively. The power of additive white Gaussian noise is assumed to be $\sigma_1^2=\sigma_2^2=\sigma_3^2=-90\text{dBm}$ and the power budgets of the BS and user A are $P_{BS} = 40\text{dBm}$ and $P_{UA} = 20\text{dBm}$, respectively. The maximum power output of the PA is set to $30\text{dBm}$ and we assume that the bandwidth of the three links are $B_{1}=B_{2}=B_{3}=100$MHz~\cite{mmWave_bandwidth}. The number of computation tasks is $L=10^6$ bits and the compression ratio is chosen as $\alpha=0.01$, which is a typical value for tasks like such as MPEG4 video (2D data and point cloud data) compression~\cite{vr_compression}. The computation capacities of the local CPU and the edge server are $F_L=200$Mbps and $F_E=1600$Mbps, respectively. The number of frames in a channel statics coherence time and the number of time slots in a frame is set to $T_f =100$ and $T_s =100$, respectively.  As for the algorithm parameters, the long-term analog beamforming design algorithm is updated based on $\varepsilon^t = 0.6^t$ and $\gamma^t = 0.9^t$. For the short-term digital beamforming design, the tolerance of accuracy is set as $\delta_1=10^{-3}$, and $\delta_2=10^{-8}$. The initial value of the penalty coefficient is $\varrho^0 = 0.1$ and the control parameter is $c=0.8$.

We first study the convergence behavior of the proposed algorithm. Fig. \ref{longterm_convergence} presents the convergence performance of the proposed long-term analog beamforming design \textbf{Algorithm 1}. The left y axis shows the value of the weighted ergodic channel capacity, which converges rapidly within 100 iterations. We also provide the value of the system latency, which is shown along the right y axis. As we can see, the system latency converges almost synchronously with the weighted ergodic capacity and decreases significantly when the iteration number increases, which validates the effectiveness of convergence for the proposed long-term algorithm. Fig. \ref{fig:fig3} presents the convergence performance of the proposed short-term digital beamforming design \textbf{Algorithm 2}. As shown in Fig. \ref{fig:fig3}\subref{fig:3a}, the objective function converges within about 40 iterations. Fig. \ref{fig:fig3}\subref{fig:3b} shows the penalty terms versus the number of iterations, which finally decreases to a level less than $10^{-12}$, indicating that the constraint is satisfied at the convergence point, thereby verifying the effectiveness of the proposed penalty-CCCP algorithm for handling the non-linear power constraint.

\begin{figure}[t]
	\centerline{\includegraphics[width=3.7in]{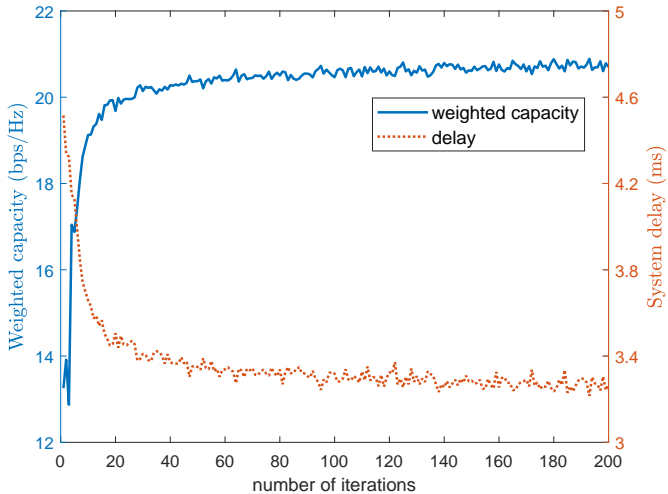}}
	\caption{Convergence performance of our proposed long-term analog beamforming design algorithm.}
	\label{longterm_convergence}
\end{figure}

\begin{figure}[t]
	\centering
	\subfloat[]{\label{fig:3a}{\includegraphics[width=0.24\textwidth]{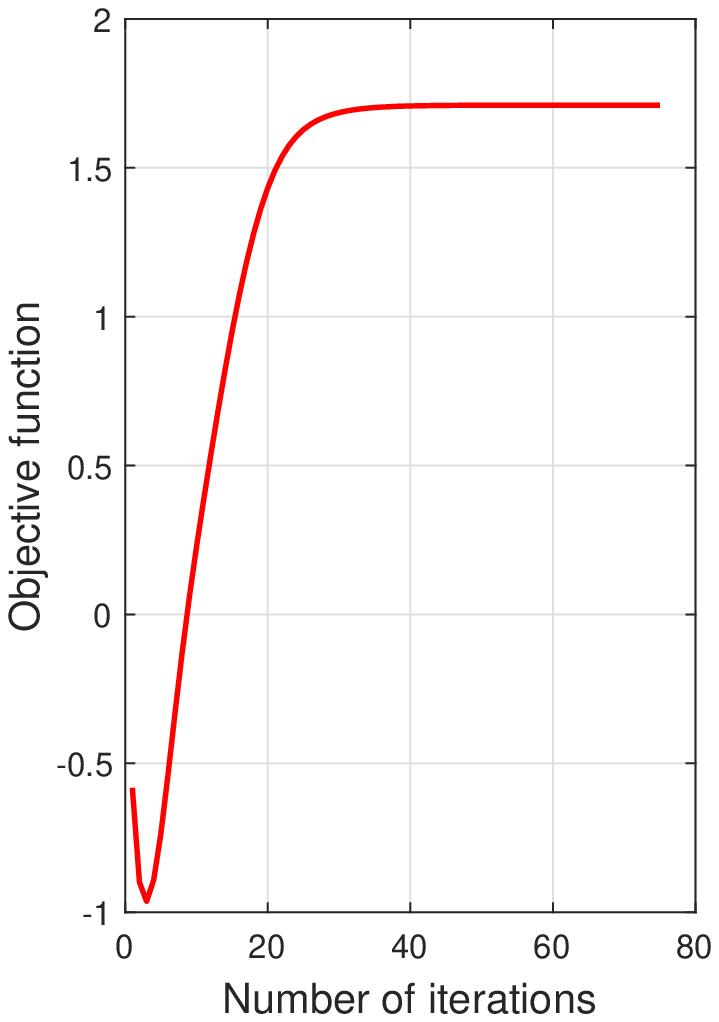}}}
	\subfloat[]{\label{fig:3b}{\includegraphics[width=0.24\textwidth]{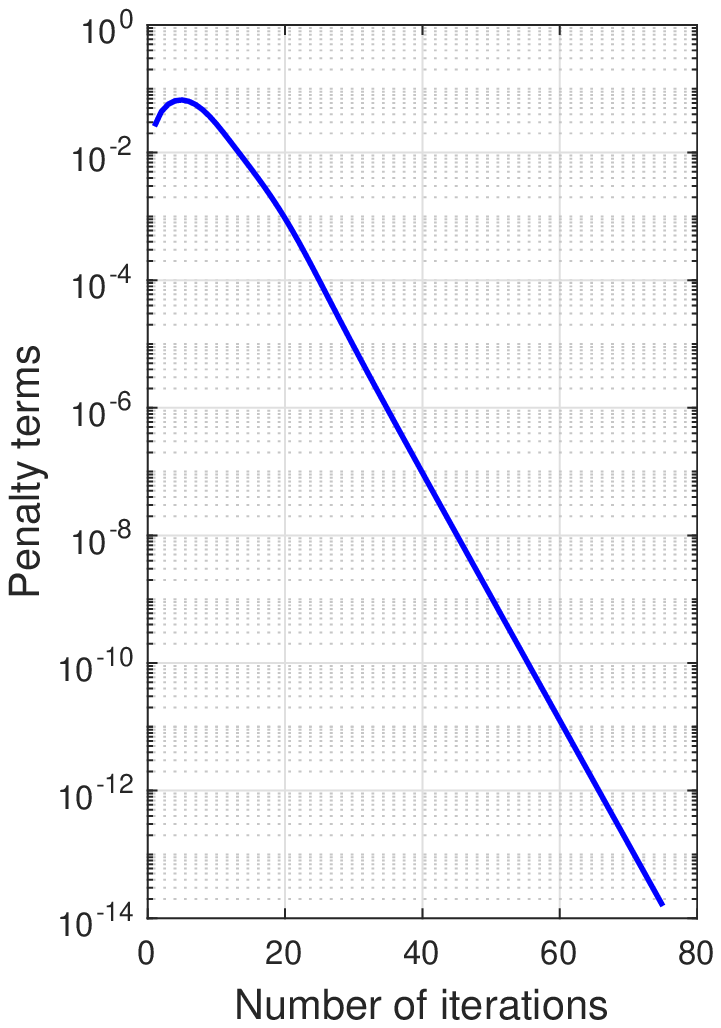}}}
	\caption{Convergence performance of our proposed short-term digital beamforming design algorithm in the uplink: (a) Objective function value versus the number of iterations; (b) Penalty terms versus the number of iterations.}
	\label{fig:fig3}
\end{figure}

In the following, we analyze the performance of our proposed two-timescale joint hybrid beamforming and offloading ratio design algorithm. We provide the following benchmarks for comparison,
\begin{itemize}
	\item Two-timescale heuristic beamforming: This scheme adopts \textbf{Algorithm 1} for the long-term analog beamforming design and the derived optimal solution for the offloading ratio design, and the proposed heuristic low-complexity algorithm is employed for the short-term digital beamforming design.
	\item Two-timescale binary offloading: This scheme adopts \textbf{Algorithm 1} for the long-term analog beamforming design and \textbf{Algorithm 2} for the short-term digital beamforming design. Moreover, the binary offloading strategy, i.e. selecting the one that has the lowest delay between the local computing scheme ($\rho = 0$) and the edge computing scheme ($\rho = 1$), is employed.
	\item Single-timescale OMP: This scheme adopts the OMP algorithm~\cite{mmWave_OMP} for the hybrid beamforming design and the optimal solution for the offloading ratio design in each time slot.
	\item Single-timescale CM: Similar to the single-timescale OMP, however, it employs the CM algorithm \cite{channelmatching} for the A/D hybrid beamforming design.
	\item Single-timescale AO: Similar to the single-timescale OMP, however, it employs the AO algorithm \cite{mmWave_AO} for the A/D hybrid beamforming design.	
\end{itemize}

%In this work, we assume that the design algorithm is implemented at the BS. Then, the required uplink CSI matrix $\mathbf{H}_1$ is estimated at the BS, and the downlink CSI matrix $\mathbf{H}_2$ and D2D CSI matrix $\mathbf{H}_3$ are estimated at user B and fed back to the BS. Hence the CSI overhead is mainly from the downlink and D2D link. The size of the CSI overhead for the two-timescale algorithm in a frame is given by $NN_b+N_aN_b+T_s[N_{rf}N_{rfb}+N_{rfa}N_{rfb}]$, while the counterpart of the single-timescale algorithm is given by $T_s[NN_b+N_aN_b]$. As we can see, the proposed two-timescale scheme can significantly reduce the required CSI overhead. Moreover, we assume that the CSI delay is proportional to the size of the required CSI matrix as \cite{CSI_delay} and set the CSI delay of the single-timescale algorithm as $\tau = 4$ms, then the CSI delay of the two-timescale algorithm can be approximately computed as $\tau_{tts} = \frac{N_{rfb}N_{rf}}{N_bN}\tau = 0.094$ms, which is significantly reduced.
We assume that the CSI delay is proportional to the size of the required CSI matrix as \cite{CSI_delay}. Specifically, the size of the CSI overhead for the two-timescale algorithm in a frame is given by $\zeta[NN_b+N_aN_b+T_s(N_{rf}N_{rfb}+N_{rfa}N_{rfb})]$, where $\zeta$ is the number of quantization bits for each element of the CSI matrices, while that of the single-timescale algorithm is given by $\zeta[T_s(NN_b+N_aN_b)]$. Fig. \ref{performance_CSI_overhead} illustrates the CSI overhead versus the number of antennas at the BS $N$. We can see that the proposed two-timescale algorithm remarkably reduces the required signaling overhead, especially when $N$ is large. In the simulation, we set the CSI delay of the single-timescale algorithm as $\tau = 4$ms. Then, the CSI delay of the two-timescale algorithm can be computed as $\tau_{tts} = \frac{N_{rfb}N_{rf}}{N_bN}\tau = 0.094$ms.
\begin{figure}[t]
	\centerline{\includegraphics[width=3.4in]{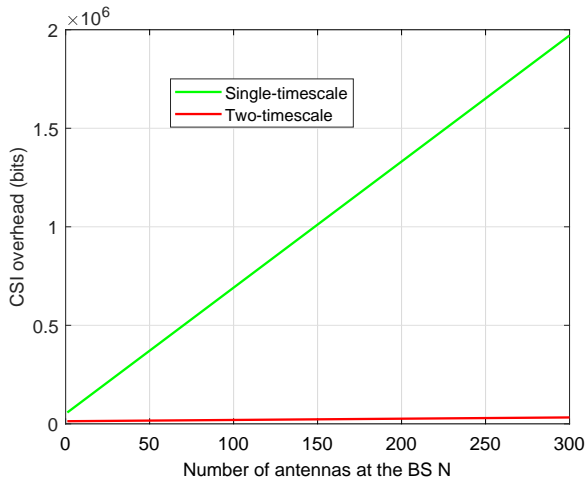}}
	\caption{CSI overhead versus number of antennas at the BS $N$.}
	\label{performance_CSI_overhead}
\end{figure}

%\begin{table}[htbp]
%	\centering
%	\caption{Size of the CSI feedback overhead versus the number of antennas}
%	\label{table2}
%	\begin{tabular}{|c|c|c|c|c|c|c|}
%		\hline
%		&10&20&30&40&50&60 \\
%		\hline
%		Single-timescale&$1.70\times10^5$&$6.80\times10^5$&$1.53\times10^6$&$2.72\times10^6$&$4.25\times10^6$&$6.12\times10^6$ \\
%		\hline
%		Two-timescale&$4.47\times10^3$&$9.57\times10^3$&$1.81\times10^4$&$3.00\times10^4$&$4.53\times10^4$&$6.40\times10^4$ \\		
%		\hline
%	\end{tabular}
%\end{table}
\begin{figure}[t]
	\centerline{\includegraphics[width=3.4in]{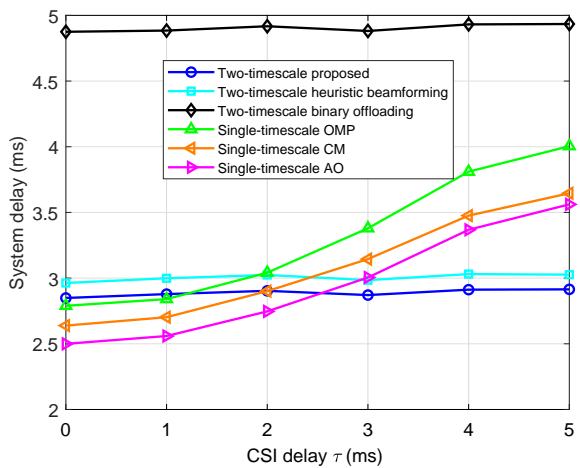}}
	\caption{System delay $T_{total}$ versus the CSI delay $\tau$ (ms).}
	\label{performance_CSI_delay}
\end{figure}

Fig. \ref{performance_CSI_delay} shows the latency performance of different algorithms versus the CSI delay. As we can see, the proposed two-timescale algorithms vary slightly when the CSI delay increases, while the conventional single-timescale algorithms degrade dramatically with the CSI delay. We also observe that the proposed algorithm outperforms the other compared algorithms when the CSI delay is larger than $3$ms. Fig. \ref{performance_transmit_power} compares the delay of different algorithms versus the transmit power of user A, i.e., $P_{UA}$. We observe that our proposed algorithm provides evident superiority over the single-timescale algorithms and the binary offloading algorithm. When the transmit power is large, the proposed heuristic short-term beamforming algorithm achieves close performance as the proposed short-term penalty-CCCP based algorithm. However, when the transmit power is small, the gap between the heuristic low-complexity algorithm and the penalty-CCCP based algorithm becomes large for the impact of the non-linear PAs becomes evident.
\begin{figure}[t]
	\centerline{\includegraphics[width=3.4in]{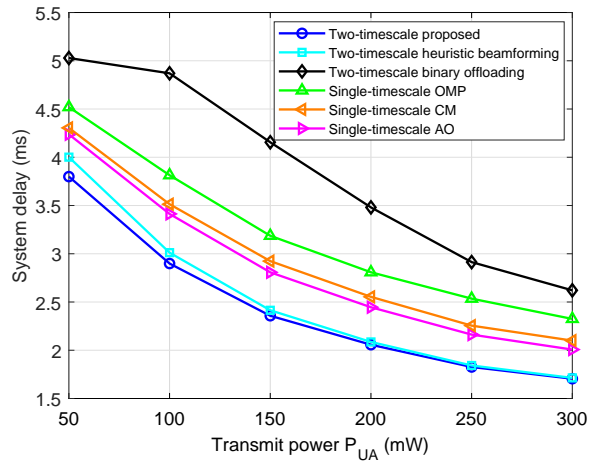}}
	\caption{System delay $T_{total}$ versus the transmit power of user A $P_{UA}$ (mW).}
	\label{performance_transmit_power} \vspace{-0.5em}
\end{figure}

Fig. \ref{performance_distance} presents the latency of various algorithms when the user position $D_y$ changes. We observe that our proposed algorithm still achieves the lowest latency performance, and as the distance between the BS and the users increases, the latency of different schemes gradually converges to the small level. This is not hard to understand because when the users are far from the BS, the users will offload less tasks to the edge server, for transmitting the raw data through the uplink is time-consuming. When the distance is quite large, the offloading ratio will be close to 0, i.e. the local computing scheme. In fact, by observing the two-timescale binary offloading algorithm, we can find that the local computing scheme starts to outperform the edge computing scheme if $D_y$ is larger than 80m.
\begin{figure}[t]
	\centerline{\includegraphics[width=3.4in]{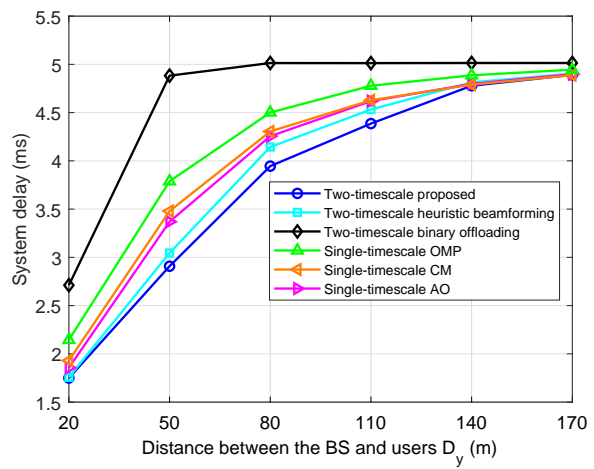}}
	\caption{System delay $T_{total}$ versus the user position $D_y$ (Mbps).}
	\label{performance_distance} \vspace{-0.5em}
\end{figure}

Fig. \ref{performance_computation_resourse} indicates the delay of different algorithms versus the computation resource ratio $\eta$, where $\eta \triangleq \frac{F_E}{F_L}$ and $F_L$ is fixed to 200MHz. We present the latency of our proposed algorithm under two transmit power settings, i.e. $P_{UA} = 100$mW and $P_{UA} = 200$mW. As we can see, when $P_{UA} = 100$mW, the proposed algorithm and the binary offloading algorithm vary slightly, even when $\eta$ is large. However, when $P_{UA} =200$mW, the proposed algorithm and the binary offloading algorithm decrease evidently with $\eta$. This is because when the system latency is limited by the transmission rate, simply increasing the computing resource will not significantly reduce the delay, which motivates us to alleviate the system bottleneck instead of simply raising the edge computing capacity.
\begin{figure}[t]
	\centerline{\includegraphics[width=3.2in]{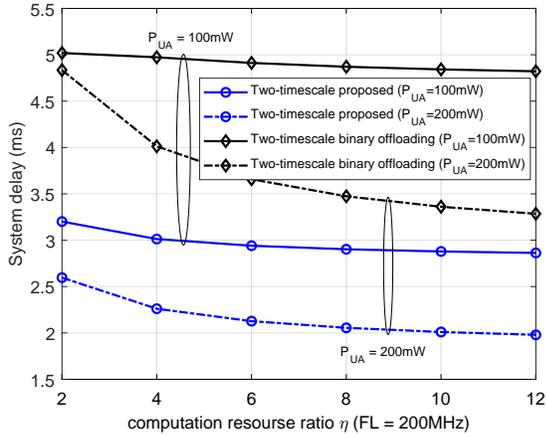}}
	\caption{System delay $T_{total}$ versus the computation resource ratio $\eta$.}
	\label{performance_computation_resourse} \vspace{-0.5em}
\end{figure}

\textcolor{black}{
Fig. \ref{performance_analog_quantization} shows the latency performance of the analyzed algorithms versus different quantization bits of the analog beamformer. We observe that latency of all algorithms decreases with the increasing of quantization bits. Moreover, the proposed algorithm only needs 4 or 5 quantization bits to achieve near performance with that of infinite quantization level, which means that the developed algorithm is efficient in practice.}
\begin{figure}[t]
\centerline{\includegraphics[width=3.34in]{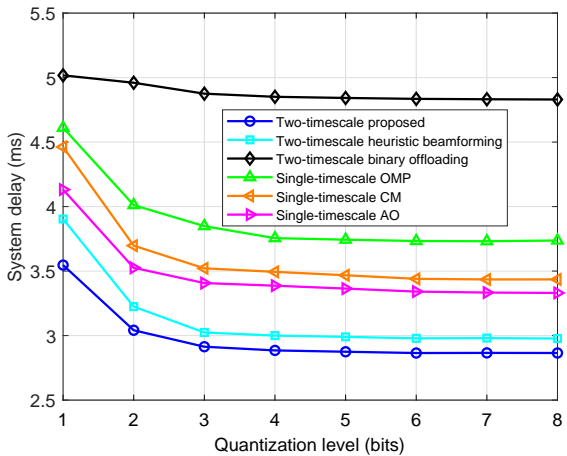}}
\caption{System delay $T_{total}$ under different quantization bits.}
\label{performance_analog_quantization} \vspace{-1em}
\end{figure}

Fig. \ref{performance_rician} illustrates the delay of different algorithms versus the Rician factor $\psi$, which is defined as $\psi = \frac{\sigma_{p1}^2}{\sum_{l=2}^{L_p}\sigma_{pl}^2}$. It is observed that when $\psi$ increases, the delay of the proposed two-timescale algorithm decreases significantly. This is because a larger Rician factor means a more deterministic channel. Hence the proposed stochastic optimization algorithm will perform better. We also observe that the latency of the single-timescale AO algorithm and the CM algorithm decreases with $\psi$, which is due to the fact that as the channel approaches rank-1, these two algorithms can find near optimal solutions. However, because of the CSI delay, our proposed two-timescale algorithm still outperforms them. The performance of the single-timescale OMP algorithm degrades severely with $\psi$  because the OMP algorithm  assumes that the LOS and NLOS components have the same gain and is not suitable for channels with large Rician factors. In contrast, our developed algorithm  does not assume a specific channel model and can be applied to a variety of channels.
\begin{figure}[t]
	\centerline{\includegraphics[width=3.4in]{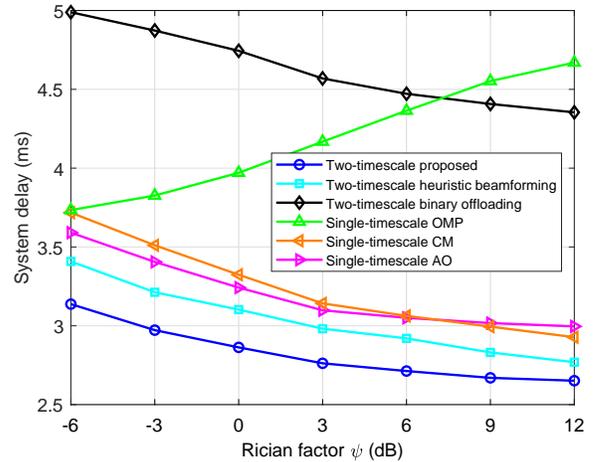}}
	\caption{System delay $T_{total}$ versus the Rician factor ($\psi$).}
	\label{performance_rician} \vspace{-1em}
\end{figure}

\vspace{-1em}
\section{Conclusion}
\label{Section7:conclusion} \vspace{-0.2em}
In this paper, we investigated a mmWave and D2D assisted MEC system, in which user A aims to process computation tasks and share the results with another user B with the aid of BS. We proposed a two-timescale algorithm where the analog beamforming matrices are updated at a long-timescale and the digital beamforming matrices and the offloading ratio are optimized at a short-timescale to reduce the required CSI overhead and minimize the system latency. We developed a SSCA-based algorithm to design the long-term analog beamforming matrices. The short-term digital beamforming matrices have been optimized relying on the concept of the penalty-CCCP for dealing with the mmWave non-linear transmit power constraint, and the offloading ratio has been obtained via the closed-form solution. We carried out the optimality and computational complexity analysis for the long-term  and short-term design algorithms, respectively. Simulation results have been provided to verify the effectiveness of our proposed joint design algorithm. Extending the mmWave and D2D assisted MEC system to the multi-user case and more specific computation model is worthy of further investigation.

\vspace{-0.7em}

\begin{appendices}
    \section{Derivation of the gradients}
    Based on the rules of matrix computation, we express the derivatives associated with the long-term analog beamforming matrices as follows \vspace{-0.5em}
    \begin{equation}
    	\frac{\partial g}{\partial \boldsymbol{\theta}_{\mathbf{U}_1}} = \frac{\partial g}{\partial \mathbf{U}_1}\circ 1j\mathbf{U}_1 - \frac{\partial g}{\partial \mathbf{U}_1^*}\circ 1j\mathbf{U}_1^*, \label{partial_phase_U1}
    \end{equation}
	\begin{equation}
		\frac{\partial g}{\partial \boldsymbol{\theta}_{\mathbf{U}_2}} = \frac{\partial g}{\partial \mathbf{U}_2}\circ 1j\mathbf{U}_2 - \frac{\partial g}{\partial \mathbf{U}_2^*}\circ 1j\mathbf{U}_2^*,
	\end{equation}
	\begin{equation}
		\frac{\partial g}{\partial \boldsymbol{\theta}_{\mathbf{F}_a}} = \frac{\partial g}{\partial \mathbf{F}_a}\circ 1j\mathbf{F}_a - \frac{\partial g}{\partial \mathbf{F}_a^*}\circ 1j\mathbf{F}_a^*,
	\end{equation}
	\begin{equation}
		\frac{\partial g}{\partial \boldsymbol{\theta}_{\mathbf{F}_b}} = \frac{\partial g}{\partial \mathbf{F}_b}\circ 1j\mathbf{F}_b - \frac{\partial g}{\partial \mathbf{F}_b^*}\circ 1j\mathbf{F}_b^*,\label{partial_phase_Fb}
	\end{equation}
	and the derivatives associated with the analog beamforming matrices are given by \vspace{-0.5em}
	\begin{equation}
		\begin{split}				
		\frac{\partial g}{\partial \mathbf{U}_1^*} =& \frac{w_1}{\sigma_1^2}[\mathbf
		{I}-\mathbf{U}_1(\mathbf{U}_1^H\mathbf{U}_1)^{-1}\mathbf{U}_1^H] \\
		&\times \mathbf{Y}_1^{-1}\mathbf{H}_1\mathbf{F}_a\mathbf{F}_a^H\mathbf{H}_1^H\mathbf{U}_1(\mathbf{U}_1^H\mathbf{U}_1)^{-1},
	\end{split}
	 \label{partial_U1}
	\end{equation}
    \begin{equation}
    	\frac{\partial g}{\partial \mathbf{U}_2^*} = \frac{w_2}{\sigma_2^2}\mathbf{H}_2^H\mathbf{F}_b(\mathbf{F}_b^H\mathbf{F}_b)^{-1}\mathbf{F}_b^H\mathbf{Y}_2^{-1}\mathbf{H}_2\mathbf{U}_2,
    \end{equation}
    \begin{equation}
    	\begin{split}    		    	
    	\frac{\partial g}{\partial \mathbf{F}_a^*} =& \frac{w_1}{\sigma_1^2}\mathbf{H}_1^H\mathbf{U}_1(\mathbf{U}_1^H\mathbf{U}_1)^{-1}\mathbf{U}_1^H\mathbf{Y}_1^{-1}\mathbf{H}_1\mathbf{U}_1 \\
    	&+\frac{w_3}{\sigma_3^2}\mathbf{H}_3^H\mathbf{F}_b(\mathbf{F}_b^H\mathbf{F}_b)^{-1}\mathbf{F}_b^H\mathbf{Y}_3^{-1}\mathbf{H}_3\mathbf{F}_b,
    \end{split}
    \end{equation}
	\begin{equation}
		\begin{split}	
	\frac{\partial g}{\partial \mathbf{F}_b^*} =& \frac{w_2}{\sigma_2^2}[\mathbf
	{I}-\mathbf{F}_b(\mathbf{F}_b^H\mathbf{F}_b)^{-1}\mathbf{F}_b^H]\\
	&\times\mathbf{Y}_2^{-1}\mathbf{H}_2\mathbf{U}_2\mathbf{U}_2^H\mathbf{H}_2^H\mathbf{F}_b(\mathbf{F}_b^H\mathbf{F}_b)^{-1} \\
	&+\frac{w_3}{\sigma_3^2}[\mathbf{I}-\mathbf{F}_b(\mathbf{F}_b^H\mathbf{F}_b)^{-1}\mathbf{F}_b^H]\\
	&\times\mathbf{Y}_3^{-1}\mathbf{H}_3\mathbf{F}_a\mathbf{F}_a^H\mathbf{H}_3^H\mathbf{F}_b(\mathbf{F}_b^H\mathbf{F}_b)^{-1},	
        \end{split} \label{partial_Fb}
    \end{equation}
    where \vspace{-0.5em}
    \begin{equation}
    	\mathbf{Y}_1 =\mathbf{I}+ \frac{1}{\sigma_1^2}\mathbf{H}_1\mathbf{F}_{a}\mathbf{F}_{a}^H\mathbf{H}_1^H\mathbf{U}_1(\mathbf{U}_1^H\mathbf{U}_1)^{-1}\mathbf{U}_1^H,
    \end{equation}
    \begin{equation}
  	\mathbf{Y}_2 = \mathbf{I}+\frac{1}{\sigma_2^2}\mathbf{H}_2\mathbf{U}_2\mathbf{U}_2^H\mathbf{H}_2^H\mathbf{F}_{b}(\mathbf{F}_{b}^H\mathbf{F}_{b})^{-1}\mathbf{F}_{b}^H,
    \end{equation}
    and \vspace{-0.5em}
    \begin{equation}
     \mathbf{Y}_3 =\mathbf{I}+\frac{1}{\sigma_3^2}\mathbf{H}_3\mathbf{F}_{a}\mathbf{F}_{a}^H\mathbf{H}_3^H\mathbf{F}_{b}(\mathbf{F}_{b}^H\mathbf{F}_{b})^{-1}\mathbf{F}_{b}^H.
    \end{equation}
    Moreover, we have $\frac{\partial g}{\partial \mathbf{U}_1} = (\frac{\partial g}{\partial \mathbf{U}_1^*})^*$, $\frac{\partial g}{\partial \mathbf{U}_2} = (\frac{\partial g}{\partial \mathbf{U}_2^*})^*$, $\frac{\partial g}{\partial \mathbf{F}_a} = (\frac{\partial g}{\partial \mathbf{F}_a^*})^*$ and $\frac{\partial g}{\partial \mathbf{F}_b} = (\frac{\partial g}{\partial \mathbf{F}_b^*})^*$ for the real value objective function. By substituting \eqref{partial_U1}$ \text{-} $\eqref{partial_Fb} into \eqref{partial_phase_U1}$ \text{-} $\eqref{partial_phase_Fb}, we finally obtain the derivatives for the long-term analog beamforming matrices.

	\vspace{0.5em}
	{\color{black}	
	\section{Derivation of updating steps in the inner loop of Algorithm 1}
	In this appendix, we provide the detailed solutions for updating the block of variables in the proposed penalty-CCCP based short-term digital beamforming algorithm.

	\textbf{Block 1}: We update $\mathbf{V}_1$ and $P_{PA1}$ in parallel with the other variables fixed. The subproblem with regard to $\mathbf{V}_1$ is given by
	\begin{equation}
		\min_{\mathbf{V}_1} \quad \text{Tr}(\mathbf{Z}\mathbf{E})
	\end{equation}
	which is an unconstrained problem. By applying the first order optimality condition, the optimal solution to $\mathbf{V}_1$ is given by
	\begin{equation}
		\mathbf{V}_1 = [\sigma_1^2\mathbf{U}_1^H\mathbf{U}_1+\mathbf{H}_{ef1}\mathbf{W}_{a1}\mathbf{W}_{a1}^H\mathbf{H}_{ef1}^H]^{-1}\mathbf{H}_{ef1}\mathbf{W}_{a1}. \label{optimal_V1}
	\end{equation}

The subproblem w.r.t. $P_{PA1}$ is given by
	\begin{subequations}
		\begin{align}
			\min_{P_{PA1}} \,\,\, & \sum_{i=1}^{N_a}(P_{PA1}(i)-h(V_{out1}(i)))^2 + (\sum_{i=1}^{N_a}P_{PA1}(i) - \bar{P}_{UA})^2 \\
			\text{s.t.} \,\,\, & \eqref{con_Ppai}. \nonumber
		\end{align}
	\end{subequations}
	This is a convex problem and the optimal solution can be expressed as
	\begin{equation}
		\begin{split}		
		P_{PA1}(i) &= \max(0,\min(4P_{max}/\pi,\\
		&(h(V_{out1}(i))+\bar{P}_{UA}-\sum_{j\neq i}^{N_a}P_{PA1}(j))/2)).
	\end{split}
	 \label{optimal_PPA}
	\end{equation}
	
	\textbf{Block 2}: We update $\mathbf{Z}$ and $V_{out1}$ in parallel by fixing the other variables. The subproblem of $\mathbf{Z}$ is given by
	\begin{equation}
		\min_{\mathbf{Z}} \text{Tr}(\mathbf{Z}\mathbf{E})-\log \det(\mathbf{Z})
	\end{equation}
	By checking the first order optimality condition, the optimal $\mathbf{Z}$ can be expressed as
	\begin{equation}
		\mathbf{Z} = \mathbf{E}^{-1} = (\mathbf{I}-\mathbf{V}_1^H\mathbf{H}_{ef1}\mathbf{W}_{a1})^{-1}, \label{optimalZ}
	\end{equation}
where the last equality holds from substituting the optimal value of $\mathbf{V}_1$, i.e. \eqref{optimal_V1} into  \eqref{E_expression}.
	
	The subproblem w.r.t. $V_{out1}$ is given by
	\begin{subequations}
		\begin{align}
			\min_{V_{out1}} \quad & \sum_{i=1}^{N_a}(P_{PA1}(i)-h(V_{out1}(i)))^2 \nonumber\\ &\qquad+\sum_{i=1}^{N_a}(\|\mathbf{F}_a(i,:)\mathbf{W}_{a1}\|-V_{out1}(i))^2 \\
			\text{s.t.} \quad & \eqref{con_Vouti}. \nonumber	
		\end{align}
	\end{subequations}
	It is readily seen that the problem can be divided into $N_a$ parallel subproblems, which yields
	\begin{subequations}
		\begin{align}
			\min_{V_{out1}(i)} \quad & \varphi(V_{out1}(i)) \triangleq (P_{PA1}(i)-h(V_{out1}(i)))^2 \nonumber \\ &+(\|\mathbf{F}_a(i,:)\mathbf{W}_{a1}\|-V_{out1}(i))^2 \\
			\text{s.t.} \quad & V_{out1}(i) \leq \sqrt{P_{max}}.	
		\end{align}
	\end{subequations}
	Since the objective function is piecewise, we need to discuss different situations and make comparison. Defining $x_1^*$ and $x_2^*$ whose expressions are given by \eqref{x_1} and \eqref{x_2}, respectively,
	\begin{figure*}[t]
	\begin{equation}
		x_1^* \triangleq \left.\frac{\pi^2(2\sqrt{P_{max}}P_{PA1}(i)/\pi +\|\mathbf{F}_a(i,:)\mathbf{W}_{a1}\|)}{\pi^2+4P_{max}}\right|_0^{\sqrt{P_{max}}/2} \label{x_1}
	\end{equation}
	\begin{equation}
		x_2^* \triangleq \left.\frac{\pi^2(6\sqrt{P_{max}}P_{PA1}(i)/\pi +\|\mathbf{F}_a(i,:)\mathbf{W}_{a1}\|+12P_{max}\sqrt{P_{max}}/\pi^2)}{\pi^2+36P_{max}}\right|_{\sqrt{P_{max}}/2}^{\sqrt{P_{max}}} \label{x_2}
	\end{equation}  \vspace{-2em}
	\end{figure*}
	where $x|_a^b \triangleq \min(\max(x,a),b)$, we can express the optimal solution to $V_{out1}(i)$ as
	\begin{equation}
		V_{out1}(i) = \begin{cases}
			x_1^*,  \qquad \varphi(x_1^*) \leq  \varphi(x_2^*), \\
			x_2^*,  \qquad \varphi(x_1^*) \geq  \varphi(x_2^*). \label{optimal_Vout}
		\end{cases}
	\end{equation}
	
	\textbf{Block 3} We update $\mathbf{W}_{a1}$ with the other variables fixed. The subproblem regarding $\mathbf{W}_{a1}$ is given by
	\begin{equation}
		\min_{\mathbf{W}_{a1}} \quad \text{Tr}(\mathbf{Z}\mathbf{E}) + \frac{1}{2\varrho}\sum_{i=1}^{N_a}(\|\mathbf{F}_a(i,:)\mathbf{W}_{a1}\|-V_{out}(i))^2. \label{object_Wa1}
	\end{equation}
	By expanding the last term of \eqref{object_Wa1} and ignoring the constant. We rewrite \eqref{object_Wa1} as
	\begin{equation}
		\min_{\mathbf{W}_{a1}} \quad \text{Tr}(\mathbf{Z}\mathbf{E}) + \frac{1}{2\varrho}\|\mathbf{F}_a\mathbf{W}_{a1}\|^2 - \sum_{i=1}^{N_a}\frac{V_{out}(i)}{\varrho}\|\mathbf{F}_a(i,:)\mathbf{W}_{a1}\|. \label{objects_Wa1}
	\end{equation}
	Note that the last term of \eqref{objects_Wa1} is concave. Hence we can approximate the original problem using the CCCP \cite{ccp}. Through the first order Taylor expansion, we provide a tight upper bound of \eqref{objects_Wa1} as follows
	\begin{equation}
		\begin{split}
		\min_{\mathbf{W}_{a1}} \quad &\text{Tr}(\mathbf{Z}\mathbf{E}) + \frac{1}{2\varrho}\|\mathbf{F}_a\mathbf{W}_{a1}\|^2 \\
		&- \sum_{i=1}^{N_a}\frac{V_{out}(i)}{\varrho}\frac{\Re\{\bar{\mathbf{W}}_{a1}^H\mathbf{F}_a(i,:)^H\mathbf{F}_a(i,:)\bar{\mathbf{W}}_{a1}\}}{\|\mathbf{F}_a(i,:)\bar{\mathbf{W}}_{a1}\|},
	\end{split}\label{upper_bound_Wa1}
	\end{equation}
	where $\bar{\mathbf{W}}_{a1}$ is the current value of variable $\mathbf{W}_{a1}$. By applying the first order optimality condition and setting $\bar{\mathbf{W}}_{a1} = \mathbf{W}_{a1}$, we express the solution to $\mathbf{W}_{a1}$ as
	\begin{equation}
		\begin{split}
			&\mathbf{W}_{a1} =  (\mathbf{H}_{ef1}^H\mathbf{V}_1\mathbf{Z}\mathbf{V}_1^H\mathbf{H}_{ef1}+\frac{1}{2\varrho}\mathbf{F}_{a}^H\mathbf{F}_a)^{-1} \\& \times(\mathbf{H}_{ef1}^H\mathbf{V}_1\mathbf{Z}+\sum_{i=1}^{N_a}\frac{V_{out1}(i)}{2\varrho \|\mathbf{F}_a(i,:)\mathbf{W}_{a1}\|}\mathbf{F}_a(i,:)^H\mathbf{F}_a(i,:)\mathbf{W}_{a1}). \label{optimal_W_a}	
		\end{split}
	\end{equation}
    }
\end{appendices}

\end{document}